\documentclass{article}
\usepackage[a4paper]{geometry}
\usepackage[pdftex]{graphicx}
\usepackage{amsmath}
\usepackage{amssymb}
\usepackage{amsthm}
\usepackage{empheq}
\usepackage{array}
\usepackage{longtable}
\usepackage{adjustbox}
\usepackage{multirow}
\usepackage{float}
\usepackage{enumitem}
\usepackage{authblk}

\usepackage{multirow}
\usepackage{subcaption}

\usepackage[hidelinks,colorlinks=true,linkcolor=blue,citecolor=blue]{hyperref}
\usepackage{xcolor}
\usepackage{tikz}
\usetikzlibrary{shapes,patterns}
\usetikzlibrary{topaths,calc}
\usetikzlibrary{decorations.pathmorphing}

\usepackage{pgfplots}
\pgfplotsset{compat=1.18}
\usepackage[ruled,vlined]{algorithm2e}
\usepackage[square,numbers]{natbib}
\usepackage{url}
\usepackage{doi}
\usepackage{appendix}
\usepackage{comment}

\title{\textbf{CSBA-Net: Learning Cross-Scale Structural Dependencies for Influential Node Identification in Complex Networks}}

\author{
\begin{center}
Rihab Boukheche$^{1}$, 
Mohammed Lalou$^{2}$, 
Nardjes Bouchemal$^{3}$ \\
$^{1}$LIRE Laboratory of Constantine 2, 
University of Mila, Algeria\\ 
\textit{rihab.boukheche@centre-univ-mila.dz}\\
$^{2}$LIB UR 7534, F-21000 Dijon, Université Bourgogne Europe, France\\
\textit{mohammed.lalou@u-bourgogne.fr}\\
$^{3}$LISI laboratory of Mila,  
University of Mila, Algeria\\
n.bouchemal@centre-univ-mila.dz
\end{center}
}

\begin{document}
\maketitle
\begin{abstract}

Identifying influential nodes is a fundamental task in complex network analysis, with applications in epidemic control, information diffusion, and network robustness. Existing centrality-based and learning-based methods often provide limited integration of structural information across different network scales. To address this limitation, we propose \textit{CSBA-Net}, for \textit{Cross-Scale Structural BiGRU Attention Network}, a multi-scale structural representation learning framework that jointly captures local, community-level, and global structural patterns. The extracted representations are processed by CNNs and integrated through a BiGRU with multi-head self-attention to model dependencies across structural scales. Influential-node scores are learned through regression using SIR-based diffusion simulations as reference targets. We conduct experiments on 24 real-world and 12 synthetic networks from social, biological, technological, and knowledge domains, and show that CSBA-Net achieves Kendall’s $\tau=0.802$, Spearman’s $\rho=0.943$, NDCG@10 $=0.898$, and NDCG@20 $=0.916$, outperforming the considered centrality-, CNN-, and GNN-based baselines. The results demonstrate the effectiveness of combining complementary structural scales for accurate influential-node ranking across diverse network structures.

\end{abstract}

\section{Introduction}
\label{sec1}

Many real-world systems can be modeled as complex network, including biological systems, transportation infrastructures, electrical power grids, and social networks. Each node within these networks is connected to many other nodes, and  a small subset of them has a crucial influence over the structural stability and functional efficiency of the system. These nodes, known as \textit{influential nodes}, act as critical drivers of different phenomena, such as the rapid spread of epidemics \cite{GUO2021127282, mata2021overview}, cascading failures across interdependent systems \cite{dong2019impact,smith2020cascading}, and the accelerated dissemination of rumors \cite{qiu2016effects,kimura2010extracting}.
Their failure or removal can compromise, or even paralyze, the entire system, potentially triggering severe consequences. Identifying them is therefore crucial to understanding system behavior.

In the literature, much research has focused on traditional centrality metrics to quantify how important a node is within a network, including degree centrality \cite{6244633}, closeness, harmonic, betweenness, and eigenvector centralities \cite{okamoto2008ranking,marchiori2000harmony,brandes2001faster,bonacich2007some}, k-shell and coreness centrality \cite{kitsak2010identification}, gateway score \cite{ruiz2014gateway}, mixed-degree decomposition \cite{ZENG20131031}, and entropy-based approaches \cite{wu2023identify}. Despite their effectiveness in providing structural insights, these approaches often fail to accurately identify node influence in complex or heterogeneous networks because they emphasize either local or global structural characteristics independently. They may also yield unreliable results in the presence of isolated or sparsely connected nodes. Moreover, nodes with similar structural features can be assigned indistinguishable importance scores, thereby reducing the discriminative capability of these measures.

More recently, alternative approaches have emerged using machine learning and deep learning techniques to address these limitations. Many machine learning approaches for node ranking have been proposed, such as Support Vector Machines \cite{wen2018fast}, K-Nearest Neighbor \cite{zhao2020machine}, Gradient Boosting Decision Trees \cite{zhang2022new}, etc. They enhance modeling flexibility by capturing complex and nonlinear relationships between features, but they still depend on handcrafted feature selection and extraction, which restricts their generalization across different network types. To overcome these limitations and better capture complex cross-scale dependencies by learning representations directly from the network structure without human intervention, several deep learning–based approaches have been developed, including classical Graph Convolutional Networks (GCNs) \cite{chen2018fastgcn,zhao2020infgcn, huang2021representation, bhattacharya2023detecting}, hybrid models integrating CNNs and GNNs \cite{zhang2022new, ramadhan2025lightweight}, Graph Attention Networks (GATs) \cite{kou2023identify}, and Transformer-based architectures \cite{jiang2025identifying}. 

%Deep learning methods, in particular graph-based neural models such as Graph Neural Networks (GNNs) \cite{chen2018fastgcn, zhao2020infgcn}, Convolutional Neural Networks (CNNs) \cite{yu2020identifying}, and Graph Convolutional Networks (GCNs) \cite{hamilton2017representation}, have demonstrated promising performance by learning representations directly from network structure without extensive human intervention. 

 An analysis of existing deep learning-based work highlights the following main limitations and challenges that remain unaddressed. First, most models formulate influential node identification as a binary classification problem, which fails to capture the continuous and heterogeneous nature of node influence. Second, standard Graph Convolutional Networks (GCNs) assign equal importance (weight) to neighboring nodes, which can reduce their effectiveness in heterogeneous or multi-community networks. Finally, many existing deep learning approaches rely on limited node features, single-scale representations, or fixed feature constructions, particularly in CNN-based methods. This restricts their ability to capture and model multi-scale structural dependencies underlying node influence in complex networks.

The need for a more flexible architecture capable of effectively integrating heterogeneous and multi-scale structural features is well motivated. To this end, we propose \textit{CSBA-Net}, for \textit{Cross-Scale Structural BiGRU Attention Network}, a novel deep learning framework to identify influential nodes in complex networks that explicitly models cross-scale dependencies using a sequential architecture, addressing the above challenges through the following contributions:

\begin{itemize}[noitemsep,label=--]
\item A new \textit{multi-scale structured representation} approach, in which each node is described through complementary micro-, meso-, and macro-level perspectives, enabling the modeling of hierarchical structural dependencies.

\item \textit{Structured neighborhood matrices} that integrate intrinsic node features with first- and second-order neighborhood information, preserving the underlying graph topology while capturing heterogeneous structural patterns across multiple scales.

\item A \textit{CNN-BiGRU architecture enhanced with multi-head self-attention}, which jointly learns hierarchical spatial structural patterns and captures inter-scale dependencies, allowing the model to adaptively focus on the most informative structural features.

\item A formulation of the problem as a \textit{continuous influence estimation task} based on SIR diffusion simulations rather than a binary classification task. The new formulation enables a more accurate and fine-grained assessment of node influence. 

 \item A \textit{network-adaptive feature balancing strategy} that integrates complementary local, community-level, and global topological descriptors within each structural matrix, preventing feature dominance and ensuring robust multi-scale representations across heterogeneous networks.

\end{itemize}

%Comprehensive experiments have been conducted on 24 real-world networks spanning social, biological, technological, and knowledge domains, as well as synthetic Barab\'{a}si--Albert benchmarks of increasing size and density, to demonstrate CSBA-Net performance and compare it with the state-of-the-art deep learning baselines in terms of ranking accuracy, robustness, and generalization.

The rest of the paper is structured as follows. In Section \ref{sec:related}, we review different approaches developed in the literature for identifying influential nodes in complex networks. Section \ref{sec:background} provides the different concepts and definitions needed in the rest of the paper. In Section \ref{sec:approach}, we describe our proposed approach to deal with the problem and detail the framework architecture. We describe the experimental protocol in Section \ref{sec:experimental} and present a comprehensive evaluation in Section \ref{sec:results}, where we discuss the results obtained on different kinds of networks. Finally, we conclude the paper and outline directions for future work.

\section{Related work}
\label{sec:related}

Several approaches have been developed in the literature to identify influential nodes in complex networks. We classify them into three main categories, namely centrality-based methods, machine learning-based methods, and deep learning-based methods. In what follows, we provide an overview of the representative works of each category and mainly focus on deep learning-based methods.

\vspace{0.15cm}
\noindent\textbf{Centrality-based methods.}
%\label{sec:first-section}  

All existing centrality-based methods leverage the topological properties of the network, and can be categorized into four main classes \cite{Lu2016vital}: \textit{(i) Neighborhood-based approaches} that evaluate the node importance based on local connectivity patterns, such as direct neighbors, or multi-step neighbors of a node for Degree and Coreness centralities \cite{6244633,kitsak2010identification}, and H-index, which gives a more nuanced view of local importance \cite{lu2016h}; \textit{(ii) Distance-based approaches}, which assess node influence  based on distances or paths within the network. This is the case for Betweenness, Closeness and Katz centralities \cite{brandes2001faster,katz1953new, okamoto2008ranking}; \textit{(iii) Iterative-based approaches}, where the node influence is continuously updated based on the importance of its neighbors, \textit{i.e.,} they recursively aggregate the influence of neighboring nodes through iterative propagation rules. This is the case for PageRank \cite{brin1998anatomy}, LeaderRank \cite{lu2011leaders}, and eigenvector centrality \cite{bonacich2007some}, and \textit{(iv) Structure-perturbation-based approaches} that determine node importance by the impact of its removal on the overall network structure and robustness \cite{tan2006evaluation}.
 
These methods often face limitations in terms of performance and flexibility, particularly in complex or heterogeneous settings. They tend to assign identical importance to nodes with the same degree while overlooking structural differences, and remain primarily based on local connectivity without fully accounting for the broader global structure of the network.

\vspace{0.15cm} 

\noindent\textbf{Machine learning-based methods.} Much research has explored machine learning techniques for identifying influential nodes, introducing a data-driven perspective in which node influence is inferred from extracted features to improve prediction performance \cite{cai2018feature}. 

Several models have been applied, including Logistic Regression \cite{ hussain2021influence}, K-Nearest Neighbors \cite{zhao2020machine}, Decision Trees \cite{zhang2022new}, Support Vector Machines \cite{wen2018fast}, and Random Forests \cite{stolarski2024identifying}. A key distinction among many of these approaches lies in how feature vectors are designed to capture node characteristics. 

 In \cite{zhao2020machine}, the authors used a vector comprising nine node features computed based on different centrality methods, and then employed seven machine learning algorithms to capture the relationship between each node’s true propagation capability and its structural characteristics. However, the dependency on manually crafted features makes it difficult to identify latent structural patterns. Similarly, Ramírez-de-la-Rosa \textit{et al.} \cite{ramirez2014towards} identified Twitter opinion leaders through machine learning algorithms by examining users’ writing styles and behaviors. Although the approach captures richer features than structural centrality alone, its reliance on textual and behavioral data limits its applicability to networks where such information is unavailable.

Wen \textit{et al.} \cite{wen2018fast} proposed a node importance evaluation method based on Least Squares Support Vector Machine (LS-SVM). Four complex indicators and an Analytic Hierarchy Process (AHP) are used to derive importance scores, which the LS-SVM then approximates to reduce computational cost. However, its performance remains bounded by the quality of the AHP scores, and the reliance on a fixed set of manually selected indicators limits generalization to different network structures and diffusion processes. In \cite{rezaei2023machine}, a Support Vector Regression (SVR) with an RBF kernel is trained on only 0.5\% of nodes selected via cluster sampling, combining degree, extended coreness and connectivity as features. While reducing training data, the method still requires complete global network information and its handcrafted features may fail to capture higher-order dependencies in heterogeneous networks.

Despite generally outperforming centrality measures, ML methods remain limited by handcrafted features, motivating representation learning, where node embeddings are learned directly from topology.

% These limitations motivate a shift toward representation learning, where node embeddings are learned directly from network topology rather than explicitly engineered.
% Overall, ML-based algorithms outperform traditional centrality measures but share a common weakness: reliance on handcrafted features designed for specific network types, making them fragile on large-scale or dynamic networks. This motivates a shift toward representation learning, where embeddings are learned directly from topology.

\vspace{0.15cm}
\noindent\textbf{Deep learning-based methods.} To address the limitations of traditional machine learning and better handle graph-structured data, several deep learning-based approaches have been introduced, including classical Graph Convolutional Networks (GCNs) \cite{zhao2020infgcn,bhattacharya2023detecting,wu2024graph,patel2025graph}, CNN-based and hybrid CNN-GNN models \cite{YU2020105893,OU2022117515,ahmad2024learning,zhang2022new,ramadhan2025lightweight}, Graph Attention Networks (GATs) \cite{kou2023identify,huang2021representation}, and Transformer-based architectures \cite{jiang2025identifying,tang2024gcnt}. 
 
For \textit{GCN-based approaches,} Zhao \textit{et al.} \cite{zhao2020infgcn} proposed  InfGCN, which leverages four node features along with neighborhood graph information. Its reliance on a fixed and limited set of features restricts its ability to capture structural patterns beyond immediate neighborhoods, and it formulates the problem as binary classification rather than continuous influence estimation. Similarly, Bhattacharya \textit{et al.} \cite{bhattacharya2023detecting} introduced DeepInfNode, which integrates degree, betweenness, k-shell, and density to learn influence representations and improve prediction performance. This model also depends on a small predefined set of topological metrics and retains the same binary classification formulation. This limits the model's ability to capture continuous and multi-level structural variation.
 
More recently, Wu \textit{et al.} \cite{wu2024graph} constructed features from regular equivalence-based similarity with adaptive eigenvalue-based selection, whereas Patel and Singh \cite{patel2025graph} derived features from degree sequences across $k$-hop neighborhoods within a modified GCN with a regression output. However, the former relies on a single global similarity relation, and the latter aggregates multi-hop information into a single homogeneous signal, rather than explicitly separating local, meso-, and global-scale structure.
 %Overall, existing GCN-based approaches either rely on predefined structural features or combine multi-hop information without explicitly separating local, meso-, and global-scale structural characteristics.

 \textit{CNN-based approaches.} Yu \textit{et al.} \cite{YU2020105893} proposed the \textit{RCNN} model, which constructs a fixed-size neighbor subgraph for each target node via breadth-first search prioritizing higher-degree neighbors. The resulting feature matrix is processed by convolutional layers to output an influence score. However, relying solely on degree to construct neighborhoods may cause peripherally located nodes to be misidentified as influential. Ou \textit{et al.} \cite{OU2022117515} extended this approach in \textit{M-RCNN} by constructing three structured matrices to capture micro-, community-, and macro-level structural information, based on neighbors' degrees, community connectivity, and k-core values, respectively. These matrices are stacked as a three-channel input and processed by a shared CNN, allowing the relative contributions of the three structural levels to be learned automatically. Note that \textit{M-RCNN} treats these levels as a fused input and does not explicitly model their sequential or hierarchical dependencies. Ahmad \textit{et al.} \cite{ahmad2024learning} proposed \textit{LCNN}, which constructs a fixed-size adjacency matrix from the target node's one-hop neighborhood. Neighbors are ranked using recursively aggregated degree information, while degree- and H-index-based multi-scale features are embedded into six channels and processed by a CNN to predict node influence. Although \textit{LCNN} reduces computational cost, its explicit neighborhood representation remains local and does not directly model mesoscopic or global structural dependencies.

  Overall, existing GCN-based approaches either rely on predefined structural features or aggregate multi-hop information into a single signal, while CNN-based methods depend on handcrafted features and encode structural information as a flat, non-hierarchical input. Consequently, both families fail to explicitly model dependencies across local, meso-, and global scales.

% Overall, existing GCN-based approaches either rely on predefined structural features or combine multi-hop information without explicitly separating local, meso-, and global-scale structural characteristics. While CNN-based methods rely on handcrafted features and represent structural information in a flat; non-hierarchical input, failing to capture dependencies across multiple network scales; meso- and macro-level.

\textit{Hybrid models} combining GNNs with other architectures, particularly CNNs, have also been explored. Zhang \textit{et al.} \cite{zhang2022new} proposed a hybrid CNN--GNN model that generates node feature matrices through a contraction-based operation over first- and second-order neighbors, avoiding predefined network metrics. However, its fixed two-hop operation yields a single-scale representation that cannot capture long-range or community-level dependencies. Ramadhan \textit{et al.} \cite{ramadhan2025lightweight} proposed a lightweight model combining 1D convolution with a GraphSAGE layer, using only degree and average neighbor degree as node features, which reduces computational complexity. These hybrid models reduce reliance on handcrafted centrality features but still represent node structure at a single-scale representation, without explicitly capturing distinct local, meso-, and global-scale structural dependencies.

 \textit{Attention and Transformer-based approaches.}
To address the limitations of fixed neighbor weighting in standard GCNs, Kou \textit{et al.} \cite{kou2023identify} proposed a graph multi-head attention regression model (GMHAR) that replaces binary classification with continuous influence prediction. Its multi-head attention mechanism adaptively weights neighboring nodes, while dense connections enhance feature propagation. However, the model relies solely on adjacency information without incorporating structural or node attributes, which may limit its generalization across heterogeneous networks.  Similarly, Huang \textit{et al.} \cite{huang2021representation} employed two Transformer-based encoders to capture structural and semantic information, which are integrated through co-attention. The model's reliance on pre-trained language models and multiple loss terms increases computational cost, and its design for attributed knowledge graphs limits direct applicability to plain complex networks. Overall, these attention-based approaches overcome fixed neighbor weighting but remain limited by their reliance on a single representation or a specific structural-semantic fusion, without explicitly modeling local, mesoscopic, and global structural roles.

 \textit{Transformer-based approaches.} Jiang \textit{et al.} \cite{jiang2025identifying} fused local and global structural information through multi-scale features but relied on manually designed metrics and added computational complexity. Tang \textit{et al.} \cite{tang2024gcnt} integrated GCNs with Graph Transformers for influence maximization, but treated the task as seed-set selection rather than node influence estimation, and its quadratic complexity limited scalability. Neither explicitly models dependencies among local, mesoscopic, and global structural roles.

In contrast, our framework explicitly models cross-scale dependencies. In what follows, we show how this is achieved by organizing micro-, meso-, and macro-level representations into a structured sequence processed by a BiGRU, while structured matrices preserve topology for efficient convolutional processing.

\section{Background and preliminaries}
 \label{sec:background} 

In this section, we introduce the fundamental concepts underlying the proposed approach.%, including complex networks, convolutional neural networks, gated recurrent units, and multi-head self-attention.

\vspace{0.1cm}
\noindent\textit{Complex networks.} A complex network can be represented as a graph $G=(V,E)$, where $V$ denotes the set of nodes and $E$ the set of edges describing their relationships. Unlike regular graphs, complex networks typically exhibit non-trivial structural properties, such as heterogeneous degree distributions, community structure, small-world characteristics, and high clustering \cite{newman2003structure,boccaletti2006complex}. In this context, influential node identification aims to estimate the relative influence of individual nodes based on their structural role and contribution to network-wide propagation.

\vspace{0.1cm}

\noindent\textit{Convolutional Neural Networks.} CNNs are deep neural architectures that use learnable convolutional filters and parameter sharing to extract local patterns from structured inputs \cite{lecun2002gradient, goodfellow2016deep}. Although originally developed for grid-structured data such as images \cite{krizhevsky2012imagenet}, CNNs can also process structured matrix representations of network neighborhoods, making them suitable for extracting local topological patterns.

\vspace{0.1cm}
\noindent\textit{Gated Recurrent Units.} GRU networks \cite{cho2014properties} use update and reset gates to regulate the integration of past and new information, enabling effective sequence modeling with relatively low computational complexity compared to LSTM \cite{chung2014empirical}. Beyond temporal data, GRUs can model interactions within structured and ordered feature representations \cite{li2015gated}. This makes them suitable for capturing dependencies among multi-scale structural representations in complex networks. In this work, a bidirectional GRU (BiGRU) is employed to model the ordered micro-, meso-, and macro-level representations in both directions, enabling information exchange across structural scales.

\vspace{0.1cm}
\noindent\textit{Multi-head self-attention.} Self-attention enables a model to assign different importance to elements of an input representation according to their pairwise relevance \cite{vaswani2017attention}. Given an input representation $Z$, the query ($Q$), key ($K$), and value ($V$) matrices are obtained as

$$
% Q=HW_Q,\qquad K=HW_K,\qquad V=HW_V,
Q=ZW_Q,\qquad K=ZW_K,\qquad V=ZW_V,
$$
 
where $W_Q$, $W_K$, and $W_V$ are learnable projection matrices that transform $Z$ into $Q$, $K$, and $V$, enabling the model to learn relevance across different structural levels.

The scaled dot-product attention is defined as

$$
\operatorname{Attention}(Q,K,V)
=
\operatorname{Softmax}
\left(
\frac{QK^\top}{\sqrt{d_k}}
\right)V.
$$

where $d_k$ is the dimensionality of the key vectors, and $\operatorname{Softmax}$ produces normalized attention weights that indicate the relative importance of each element. %The resulting attention output is a weighted combination of the value vectors, where weights are determined by the similarity between queries and keys. 

Multi-head attention performs this operation across multiple projection subspaces, allowing the model to capture complementary interactions among input representations \cite{vaswani2017attention}. In the proposed model, multi-head self-attention is applied to the BiGRU outputs to adaptively capture dependencies among different structural levels.

\section{Proposed Framework}
\label{sec:approach}

\subsection{Overview} 

\begin{figure*}[!h]
    \centering
    \includegraphics[width=0.8\textwidth]{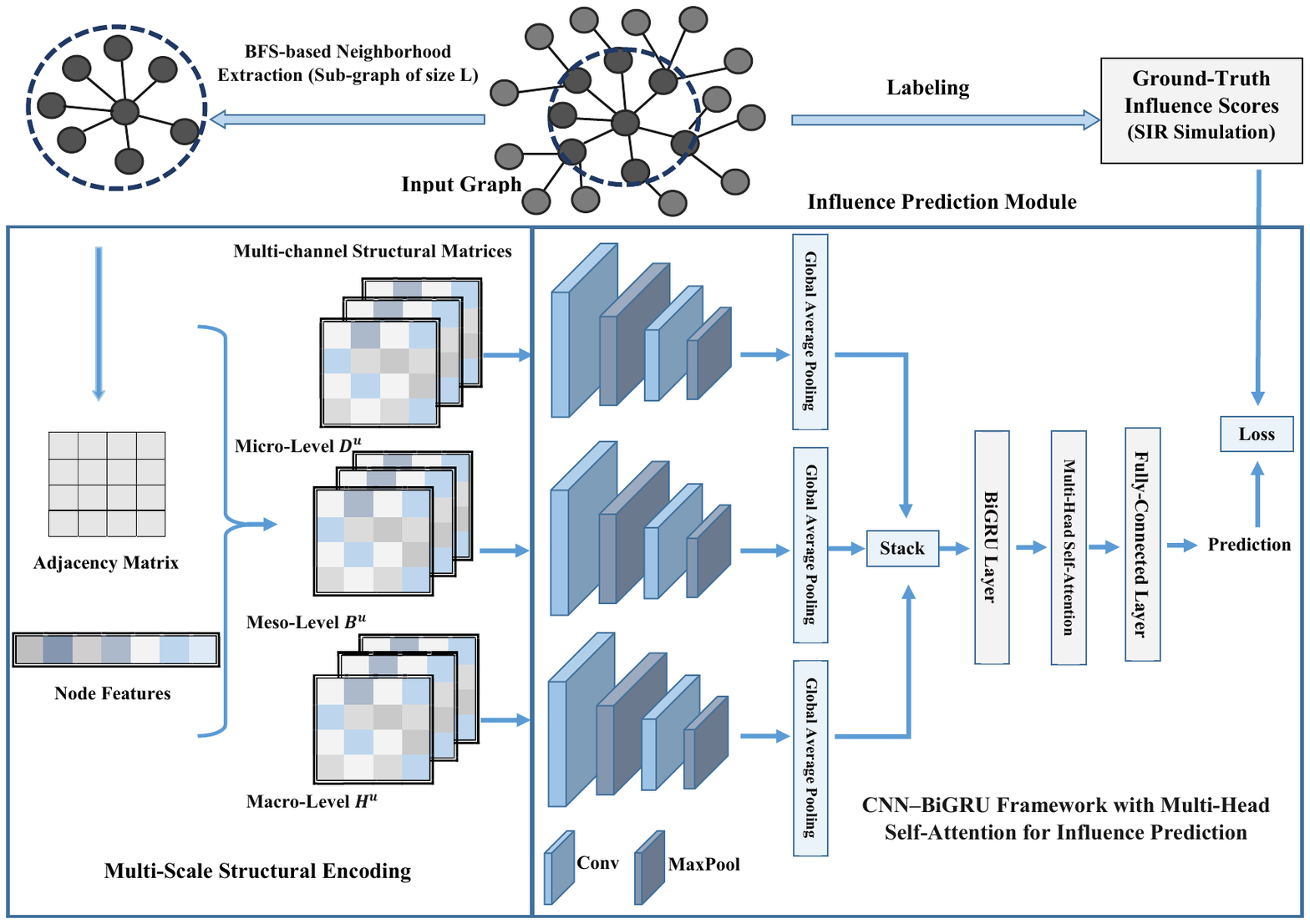}
\caption{%Schematic overview of the proposed CNN--BiGRU framework with multi-head self-attention for influential node identification. A fixed-size neighborhood subgraph is first extracted using a BFS-based strategy and encoded into multi-channel structured matrices corresponding to micro-level ($D^u$), meso-level ($B^u$), and macro-level ($H^u$) structural representations. Each matrix is processed by a dedicated CNN branch to learn hierarchical spatial features, followed by global average pooling. The resulting multi-scale features are then fused and passed to a BiGRU layer to model inter-level feature dependencies, while a multi-head self-attention mechanism selectively emphasizes the most informative structural patterns. The final fully connected layers output the predicted continuous influence score, supervised by SIR-based diffusion labels.
\textit{The proposed CNN--BiGRU framework with multi-head self-attention for influential node identification. BFS-based neighborhood subgraphs are encoded into micro-, meso-, and macro-level matrices (\(D^u\), \(B^u\), \(H^u\)), processed by parallel CNN branches, and fused for BiGRU-based inter-level modeling and attention-based feature selection. Fully connected layers then predict influence scores supervised by SIR-based labels.}
}

    \label{fig:model_propose}
\end{figure*}

The proposed framework \textit{CSBA-Net} aims to identify influential nodes in complex networks by learning hierarchical structural dependencies through a multi-stage deep learning architecture (see \textit{Figure~\ref{fig:model_propose}} for a detailed illustration of its architecture).

Given a graph $G=(V,E)$, each node $u\in V$ is represented through multi-scale structural information extracted from its local neighborhood. Specifically, a fixed-size neighborhood subgraph is extracted for each node and encoded into structured representations that capture complementary structural properties at different levels of abstraction. These representations are then encoded by parallel Convolutional Neural Network (CNN) modules that extract spatial structural patterns from each scale. The resulting embeddings are subsequently organized as a sequence and processed by a BiGRU to capture cross-scale dependencies. To further refine the learned representations, a multi-head self-attention mechanism is applied to selectively emphasize the most informative structural features. Finally, the aggregated representation is passed through fully connected layers to predict a continuous influence score for each node. Unlike binary classification approaches, CSBA-Net formulates influential node identification as a regression problem, where node influence is modeled as a continuous value. The ground-truth labels are generated using SIR diffusion simulations, enabling a more realistic characterization of node influence.

This design allows the model to jointly capture local, community-level, and global structural information, while explicitly modeling interactions across these scales in a unified learning framework. In what follows, we detail and motivate each element of the CSBA-Net architecture.

\subsection{Multi-Scale Structural Representation} 
First, to enable this framework, we construct a meaningful structural representation for each node using a multi-scale approach. For each node, we extract the neighborhood subgraph and then transform it into learnable multi-scale feature representations through the construction of structured feature matrices.

% First, to enable this framework, a meaningful structural representation for each node must be constructed, which is achieved through a multi-scale representation. For each node, we extract the neighborhood subgraph, and then transform it into multi-scale feature learnable representations through the construction of structured feature matrices.

\vspace{0.1cm}
\noindent\textit{Neighborhood Subgraph Extraction.} 
  
  The influence of a node is closely related to the structural properties of its surrounding neighborhood. To capture this context, we extract, for each target node $u$, a fixed-size neighborhood subgraph $G_u$ using Breadth-First Search until $L$ nodes are selected ($L$ is a fixed subgraph size, which is essential for efficient mini-batch training). Candidate neighbors are ranked within each BFS level by the sum of their min-max normalized betweenness and harmonic centralities to prioritize structurally important and diffusion-relevant nodes rather than relying only on degree (which may bias toward highly connected hubs, as in many existing CNN-based models).
  % Betweenness centrality favors structural bridges, whereas harmonic centrality captures distance-based reachability and remains well-defined on disconnected graphs.  

In the case of nodes sharing identical ranking scores, selection is performed randomly to avoid deterministic bias. Also, if fewer than $L-1$ neighbors are available, zero-padding is applied to maintain a fixed-size representation. The adjacency matrix $A_u$, which represents the resulting subgraph $G_u$, preserves the local topology around $u$ and serves as the basis for constructing micro-, meso-, and macro-level structural representations.

%We note that matrix $A_u$ serves as the foundation for constructing multi-scale structured representations that capture micro-, meso-, and macro-level structural dependencies. Thus, once the neighborhood subgraph is extracted, the next step consists in transforming this structural information into multi-scale feature learnable representations through the construction of structured feature matrices.
 
\vspace{0.1cm}
\noindent\textit{Structural Feature Construction.} To jointly capture intrinsic node properties and neighborhood structure, each target node $u$ is represented by three structured matrices, $D^u$, $B^u$, and $H^u$, encoding micro-, meso-, and macro-level topology, respectively. The three matrices use the same three-channel aggregation scheme: the first channel $X^{(1)}$ represents the target node through its intrinsic features $f(u)$, while the second and third, $X^{(2)}$ and $X^{(3)}$, aggregate first- and second-order neighborhood features, denoted $N(u)$ and $N_2(u)$, respectively:

\[
X^{(1)}_u = f(u), \quad
X^{(2)}_u = \sum_{v \in {N}(u)} f(v), \quad
X^{(3)}_u = \sum_{w \in {N}_2(u)} f(w)
\]

Sum aggregation is used to preserve both neighborhood size and cumulative structural strength, which are key factors in assessing node influence. Note that mean aggregation can reduce structural signals by creating
an equivalence between nodes that exhibit inherently different neighborhood strengths. Also, a three-channel design is adopted to balance representational capacity and model robustness, as additional higher-order channels may increase complexity, overfitting, redundancy, and noise.

At each scale, the neighborhood adjacency matrix $A^u$ is enriched with its own structural features: degree and clustering coefficient describe local connectivity at the micro level; betweenness, gateway score, and community closeness characterize inter-community roles at the meso level; and harmonic and eigenvector centralities capture global reachability and influence at the macro level. This feature selection is designed to provide complementary structural information across the three network scales (see Table \ref{tab:structural_properties}).

\begin{table*}[t]
\centering
\caption{Structural properties used for node representation and their diffusion-aware motivations}

\renewcommand{\arraystretch}{1.3}
\resizebox{0.99\textwidth}{!}{
\begin{tabular}{|c|c|p{10cm}|c|}
\hline
\textbf{Property} & \textbf{Scope} & \textbf{Motivation} & \textbf{Normalized}\\
\hline
Degree 
& \multirow{4}{*}{Local} 
& Measures a node's immediate connectivity and reflects its direct influence potential over neighboring nodes.&Yes\\
\cline{1-1}\cline{3-4}
Clustering Coefficient 
&  
&Captures the cohesiveness of a node’s neighborhood, which affects how influence circulates within tightly connected regions. &No \\

\hline
Betweenness Centrality 
& \multirow{6}{*}{Inter-community} 
&Identifies nodes that act as intermediaries in information flow, enabling influence to propagate across different regions of the network. &Yes  \\

\cline{1-1}\cline{3-4}
Gateway Score 
& 
& Highlights nodes acting as structural bridges across communities, playing a crucial role in cross-community influence dissemination. &No\\
\cline{1-1}\cline{3-4}

Community Closeness 
& 
&  Evaluates how efficiently a node interacts within and across communities, reflecting its accessibility and communication reach. & No\\

\hline
Harmonic Centrality 
& \multirow{4}{*}{Global} 
&Captures global reachability while remaining robust to disconnected components, reflecting a node’s ability to spread influence rapidly across long distances. &Yes\\

\cline{1-1}\cline{3-4}
Eigenvector Centrality 
& 
&Measures influence by assigning higher importance to nodes connected to other influential nodes, emphasizing global propagation strength. This property is adaptively weighted by $\beta_2$ to prevent over-amplification of hubs.& No\\
\hline
\end{tabular}
}
\label{tab:structural_properties}
\end{table*}

Let $G_u=(V_u,E_u)$ be the neighborhood subgraph centered at $u$, with $V_u=\{u_0,u_1,\ldots,u_{L-1}\}$ and $u_0=u$. The micro-level matrix $D^u$ is defined as:

\begin{equation}
\underset{\forall i,j \in \{0,\dots,L-1\}}{D^u[i,j]} =
\begin{cases}
A^u[i,j]\big(\mu_1 D_{u_j} + \mu_2 C_{u_j}\big)
& i = 0,\, j \neq 0 \\
A^u[i,j]\big(\mu_1 D_{u_i} + \mu_2 C_{u_i}\big)
& i \neq 0,\, j = 0 \\
D_{u_i} & i = j \\
A^u[i,j] & \text{otherwise}.
\end{cases}
\label{eq:micro_matrix}
\end{equation}

where $D_{u_i}$ and $C_{u_i}$ denote degree and clustering coefficient of node $u_i$, respectively, and the weighting coefficients $\mu_1,\mu_2$ control their relative contributions.

At the meso level, $B^u$ integrates betweenness centrality ($B$), gateway score ($G$), and community closeness ($CC$):

{\fontsize{9pt}{9pt}\selectfont
\begin{equation}
B^u[i,j] =
\begin{cases}
A^u[i,j]\Big( B_{u_j}\big(1 + \omega_1 G_{u_j}\big) + \omega_2\, CC_{u_j} \Big),
& i = 0,\, j \neq 0, \\[6pt]
A^u[i,j]\Big( B_{u_i}\big(1 + \omega_1 G_{u_i}\big) + \omega_2\, CC_{u_i} \Big),
& i \neq 0,\, j = 0, \\[6pt]
G_{u_i}, & i = j, \\[4pt]
A^u[i,j], & \text{otherwise}.
\end{cases}
\label{eq:comm_matrix}
\end{equation}
}

%$\omega_1$ and $\omega_2$ control the contributions of gateway structure and community accessibility, respectively. 

Note that $(1+\omega_1G)$ enhances betweenness centrality without replacing structural mediation (gateway effects vanish when betweenness is zero). Also, $\omega_2CC$ captures intra-community accessibility, enabling locally well-positioned nodes to contribute even without bridging roles. This balances structural bridging and community influence in $(B^u)$.

At the macro level, $H^u$ combines harmonic ($H$) and eigenvector ($E$) centralities:
\begin{equation}
H^u[i,j] =
\begin{cases}
A^u[i,j]\big( \beta_1 H_{u_j} + \beta_2 E_{u_j} \big),
& i = 0,\, j \neq 0, \\[6pt]
A^u[i,j]\big( \beta_1 H_{u_i} + \beta_2 E_{u_i} \big),
& i \neq 0,\, j = 0, \\[6pt]
E_{u_i}, & i = j, \\[6pt]
A^u[i,j], & \text{otherwise}.
\end{cases}
\label{eq:macro_matrix}
\end{equation}

The coefficients $\mu_1$, $\mu_2$, $\beta_1$, $\beta_2$, $\omega_1$, and $\omega_2$ are data-driven coefficients (Section \ref{sec:hyper}) that control the relative contributions of the selected structural features and adapt the representation to each network's characteristics. In fact, by integrating these features with $A^u$, the three matrices preserve the original neighborhood topology while encoding complementary information at different structural scales.

\begin{figure*}[!h]
    \centering
    \includegraphics[width=0.9\textwidth]{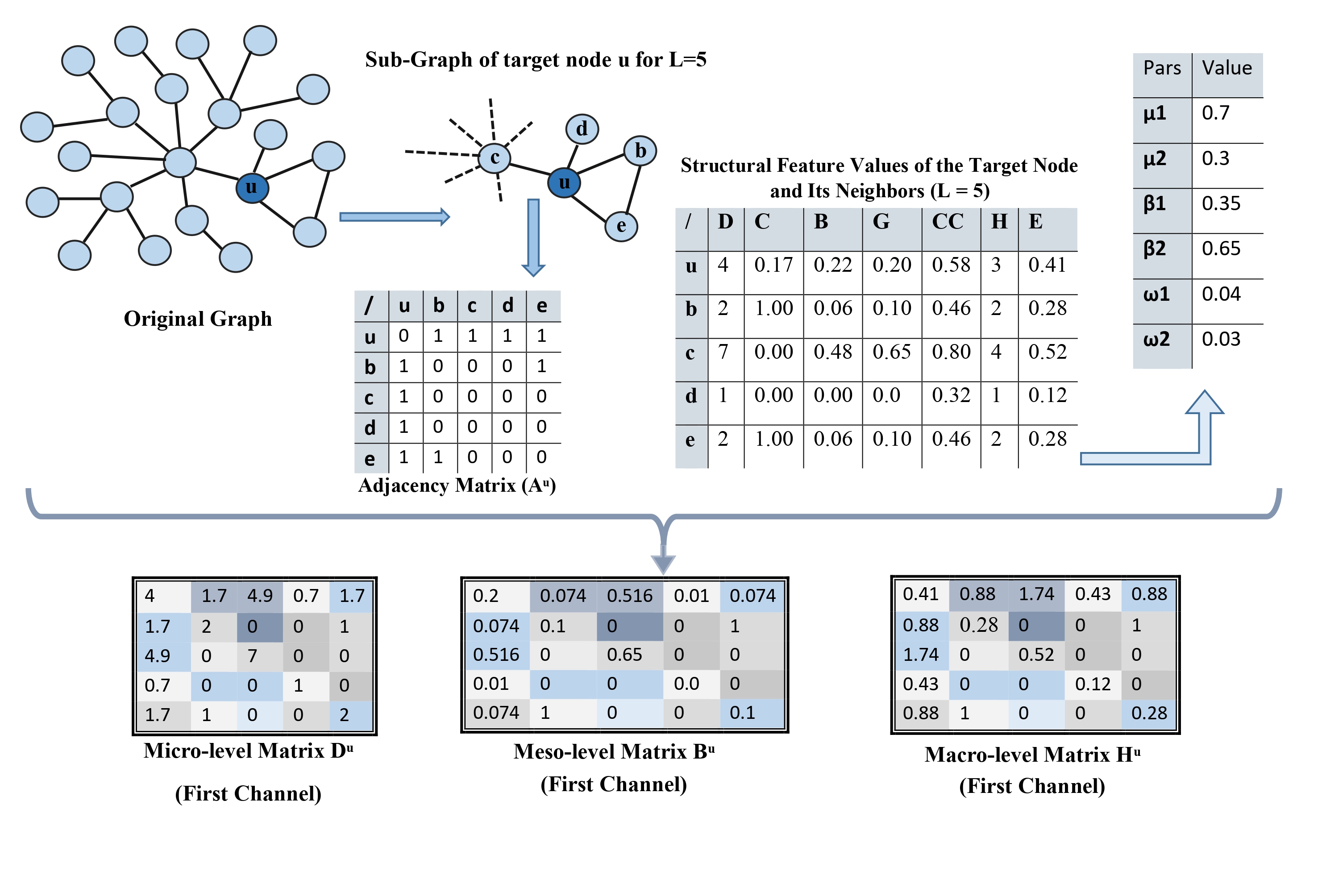}
    \caption{
    \textit{Illustrative construction of \(D^u\), \(B^u\), and \(H^u\) for target node \(u\) with \(L=5\). A BFS strategy is applied to extract the neighborhood subgraph centered at $u$, and then its adjacency matrix \(A^u\) is constructed.  The structured matrices \(D^u\), \(B^u\), and \(H^u\) are generated from graph features and the weights $\mu_1,\mu_2,\beta_1,\beta_2,\omega_1,\omega_2$. Only the first channel is shown. The feature and weight values are chosen for illustration only, to show how the matrices are computed in the actual framework.}
     }
       \label{fig:model_calcul_matrix}
\end{figure*}

Note that $D^u$, $B^u$, and $H^u$ jointly form a nine-channel representation that enriches the node embedding. After normalization and feature-specific weighting, the matrices are organized into multi-channel tensors and passed to the convolutional encoding stage for hierarchical feature extraction. An example for $L=5$ is shown in Fig.~\ref{fig:model_calcul_matrix}.

%After normalization and weighting to balance heterogeneous feature scales (hyperparameters are defined based on the statistical properties of each feature across the network), these matrices are organized as multi-channel tensors and provided to the convolutional encoding stage for hierarchical feature extraction.

    % \textit{Illustrative construction of \(D^u\), \(B^u\), and \(H^u\) for target node \(u\) with \(L=5\). A BFS strategy is applied to extract the neighborhood subgraph centered at $u$, and then its adjacency matrix \(A^u\) is constructed. The structured matrices $D^u$, $B^u$, and $H^u$ are generated from graph features and data-driven weights $\mu_1,\mu_2,\beta_1,\beta_2,\omega_1,\omega_2$. Only the first channel is shown.} 

\subsection{CNN-Based Encoding} 

Three parallel CNN branches independently process the structured matrices $D^u$, $B^u$, and $H^u$, corresponding to the micro-, meso-, and macro-level representations. Each matrix is encoded as a $3\times L\times L$ tensor, whose channels represent the target node and its first- and second-order neighbors.

Each branch consists of two convolutional layers, each followed by a LeakyReLU activation function and a \(2 \times 2\) max-pooling operation. The first layer maps the 3-channel input to 16 feature maps, while the second increases this to 32 channels. Note that both convolutions use $2\times2$ kernels with stride and padding set to 1, enabling hierarchical extraction of structural patterns within each scale. 

The LeakyReLU activation is defined as:

\[
\text{LeakyReLU}(x) =
\begin{cases}
x, & \text{if } x \ge 0, \\
\kappa x, & \text{if } x < 0,
\end{cases}
\quad \text{where } \kappa = 0.1
\]

LeakyReLU provides stable gradient propagation and avoids the dead-neuron problem of standard ReLU. The max-pooling progressively reduces the spatial dimensions while preserving the most salient structural patterns. After the two convolution–pooling stages, each branch produces a $32\times10\times10$ feature map (for L = 40, as set in Section \ref{sec:experimental}). Global average pooling then aggregates each map into a compact 32-dimensional representation, yielding $F_D,F_B,F_H\in\mathbb{R}^{32}$.

These scale-specific vectors summarize the learned structural patterns and are subsequently used for inter-scale modeling. Processing each structural level independently preserves its distinctive information while enabling complementary representations to be integrated in later stages.

%These vectors summarize the structural patterns extracted from each scale and serve as the input for subsequent inter-scale modeling. By processing each structural level independently, the CNN branches learn hierarchical spatial features specific to each scale, while preserving their complementary characteristics for later fusion.

%These multi-scale representations are subsequently used for inter-level feature modeling while preserving the complementary structural information captured by each CNN branch.

\subsection{BiGRU-based Fusion}

The CNNs extract structural patterns independently at each scale but do not capture dependencies between structural levels. To enable cross-scale modeling, their outputs are stacked into a sequence:

$$
X_{\text{seq}}=[F_D,F_B,F_H]\in\mathbb{R}^{3\times32},
$$

This allows the model to treat different structural levels as ordered elements, enabling the learning of interactions across levels. The sequence is then processed by a Bidirectional Gated Recurrent Unit (BiGRU) with hidden size $h=128$:

$$
Z=\mathrm{BiGRU}(X_{\text{seq}})\in\mathbb{R}^{3\times256},
$$

where 256 results from concatenating the forward and backward hidden states. This sequential fusion captures dependencies among micro-, meso-, and macro-level representations (rather than treating each scale independently, which is the case for most existing approaches). Since the contribution of different structural levels may vary, a multi-head self-attention mechanism is subsequently employed to adaptively emphasize the most informative representations for influence prediction.
 
\subsection{Multi-Head Self-Attention}

As not all components contribute equally to the prediction of node influence, a multi-head self-attention mechanism is applied to adaptively weight the cross-scale representations learned by the BiGRU. Following the standard formulation in Section~\ref{sec:background}, the query, key, and value matrices are obtained through learnable projections $W^Q$, $W^K$, and $W^V$, and the scaled dot-product attention is computed as

$$
O
=\mathrm{Attention}(Q,K,V)
=\mathrm{Softmax}\left(\frac{QK^{\top}}{\sqrt{d_k}}\right)V,
$$

where $d_k$ denotes the key dimension. Multiple heads operate in parallel to capture complementary interactions among the micro-, meso-, and macro-level representations. Their outputs are concatenated and linearly projected, followed by mean pooling across the sequence dimension to produce a unified embedding:

$$
c=\mathrm{Mean}(O)\in\mathbb{R}^{256},
$$

%which summarizes the most informative cross-scale dependencies and is passed to the final regression layer for influence prediction.

The resulting context vector \(c\) provides a compact representation of the most informative cross-level dependencies and is passed to the final regression layer for influence prediction.

\subsection{Influence Prediction and Training Objective} 

The final node representation $c$ is passed through two fully connected layers to predict the influence score:
\begin{equation}
\mathbb{R}^{256} \rightarrow \mathbb{R}^{1024} \rightarrow \mathbb{R}^{1}
\end{equation}
The first layer projects the representation to $1024$ dimensions using a LeakyReLU activation, and the second layer outputs a scalar value $\hat{y}_u$, representing the predicted influence of node $u$.

To train the model, a regression objective is adopted. The ground-truth influence labels are generated using the Susceptible–Infected–Recovered (SIR) diffusion model (for more details, refer to \cite{hethcote2000mathematics}). The SIR model simulates the spreading capability of each node within the network and provides a realistic approximation of its influence. For each node, multiple independent SIR simulations are performed to ensure statistical stability. The influence score of node \( u \), denoted as \( y_u \), is computed as the average number of infected and recovered nodes across all simulations, following Kitsak \textit{et al.} \cite{kitsak2010identification}:

\begin{equation}
 y_u = IA(u) = \frac{1}{T} \sum_{j=1}^{T} \left( |R_j| + |I_j| \right)
\label{eq:IA_basic}
\end{equation}

where \( R_j \) is the set of recovered nodes in the \( j \)-th simulation, and \( I_j \) is the set of infected nodes remaining at the end of the \( j \)-th simulation. This provides a realistic approximation of node influence, reflecting its ability to propagate information through the network.

The model is trained by minimizing the discrepancy between the predicted score $\hat{y}_u$ and the ground-truth score $y_u$ using the Smooth L1 loss function:

\begin{equation}
\text{SmoothL1}(e) =
\begin{cases}
\frac{e^2}{2}, & |e| < 1 \\
|e| - \frac{1}{2}, & \text{otherwise}
\end{cases}
\end{equation}
where $e = y_u - \hat{y}_u$ denotes the prediction error. This loss function combines the advantages of both L1 and L2 losses, providing robustness to outliers while maintaining sensitivity to small errors. 

%The model parameters are optimized using the Adam optimizer with a learning rate of $5 \times 10^{-4}$, a batch size of 32, and 100 training epochs. %By formulating the problem as a continuous regression task and leveraging diffusion-based labels, the proposed framework achieves a more accurate and fine-grained estimation of node influence compared to traditional classification-based approaches.

 \textit{Algorithm~\ref{alg:cnn_gru_att}} summarizes the different steps of our proposed framework. % based on CNN and BiGRU with multi-head self-attention.  

\begin{algorithm}[t]
\caption{CNN--BiGRU framework with multi-head self-attention}
\label{alg:cnn_gru_att}

\KwIn{Graph $G=(V,E)$, neighborhood size $L$, learning rate $lr$, epochs $N_{\mathrm{ep}}$ }
\KwOut{Predicted influence scores $\hat{y}_u$ for all $u\in V$}

\textbf{Preprocessing:}\

\ForEach{node $u\in V$}{
    Extract neighborhood subgraph $G_u$ of size $L$ using BFS\;
    Construct adjacency matrix $A^u$\;
    Compute structured matrices $D^u$, $B^u$, and $H^u$\;
    Encode each matrix as a three-channel tensor\;
    Generate influence label $y_u$ using SIR simulations\;
}

\textbf{Training:}\

\For{$epoch=1$ \KwTo $N_{\mathrm{ep}}$ }{
    Shuffle nodes and create mini-batches\;
    
    \ForEach{mini-batch $\mathcal{M}$}{
        \ForEach{node $u\in\mathcal{M}$}{
            $F_D \leftarrow \mathrm{CNN}(D^u)$\;
            $F_B \leftarrow \mathrm{CNN}(B^u)$\;
            $F_H \leftarrow \mathrm{CNN}(H^u)$\;
            $X_{\mathrm{seq}} \leftarrow \mathrm{Stack}(F_D,F_B,F_H)$\;
            $\mathcal{Z}_u \leftarrow \mathrm{BiGRU}(X_{\mathrm{seq}})$\; 
            $\mathcal{O}_u \leftarrow \mathrm{MultiHeadAttention}(\mathcal{Z}_u)$\;
            $c_u \leftarrow \mathrm{Mean}(\mathcal{O}_u)$\;
            $\hat{y}_u \leftarrow \mathrm{FC}(c_u)$\;
        }
        
        Compute Smooth L1 loss over $\mathcal{M}$ and backpropagate\; 
        Update parameters using the Adam optimizer with learning rate $lr$\;
        
    }
}

\Return{$\hat{y}_u$ for all $u\in V$}\;

\end{algorithm}

\subsection{Hyperparameter Definition} 
 \label{sec:hyper}
 
The hyperparameters $\mu_1,\mu_2,\beta_1,\beta_2,\omega_1,\omega_2$ are determined using a data-driven strategy that accounts for the scale and distribution of heterogeneous structural features. This calibration ensures balanced feature contributions, prevents large-magnitude features from dominating the representations, and improves their integration during learning and attention-based fusion.

At the micro level, degree centrality is normalized to reduce its dependence on network size, whereas the clustering coefficient already lies in \([0,1]\). The weights \(\mu_1\) and \(\mu_2\) are chosen to balance their average contributions:

\begin{equation}
\mu_1 \cdot \mathbb{E}[\mathrm{deg}_{\mathrm{norm}}]
\approx
\mu_2 \cdot \mathbb{E}[\mathrm{clustering}],
\quad \text{under } \mu_1 + \mu_2 = 1 \notag
\end{equation}

Solving this constraint yields: 

\begin{comment}
\begin{equation}
\begin{aligned}
\mu_1 =
\frac{\mathbb{E}[\mathrm{deg}_{\mathrm{norm}}]}
{\mathbb{E}[\mathrm{deg}_{\mathrm{norm}}] + \mathbb{E}[\mathrm{clustering}]} \\[10pt]
\mu_2 =
\frac{\mathbb{E}[\mathrm{clustering}]}
{\mathbb{E}[\mathrm{deg}_{\mathrm{norm}}] + \mathbb{E}[\mathrm{clustering}]}
\end{aligned}
\end{equation}
\end{comment}

$$
\mu_1=
\frac{\mathbb{E}[\mathrm{clustering}]}
{\mathbb{E}[\mathrm{deg}_{\mathrm{norm}}]+\mathbb{E}[\mathrm{clustering}]},
\mu_2=
\frac{\mathbb{E}[\mathrm{deg}_{\mathrm{norm}}]}
{\mathbb{E}[\mathrm{deg}_{\mathrm{norm}}]+\mathbb{E}[\mathrm{clustering}]}
$$

where \(\mathbb{E}[\cdot]\) denotes the empirical mean over all network nodes.

At the community level, \(\omega_1\) and \(\omega_2\) balance gateway effects, normalized betweenness centrality, and community-level accessibility in \(B^u\). Betweenness centrality is normalized to mitigate its skewed distribution, while gateway score and community closeness remain unnormalized to preserve their structural meaning. The gateway weight is defined as

$$
\omega_1=\frac{0.1}{\mathbb{E}[\mathrm{gateway}]}
$$

providing a moderate, data-adaptive amplification of inter-community connectivity. \(\omega_2\) balances bridging and intra-community accessibility by matching their expected contributions:

$$
\omega_2=
\frac{\mathbb{E}[\mathrm{betweenness}_{\mathrm{norm}}]}
{\mathbb{E}[\mathrm{community\_closeness}]}.
$$

%Thus, \(\omega_1\) controls gateway amplification, while \(\omega_2\) maintains a balanced contribution between inter- and intra-community structural roles.

At the macro level, harmonic centrality is normalized to account for variations in network size and path-length distributions, while eigenvector centrality remains unnormalized to preserve its global structural meaning. Following the same principle, \(\beta_1\) and \(\beta_2\) satisfy

\begin{comment}
\begin{equation}
\begin{aligned}
\beta_1=
\frac{\mathbb{E}[\mathrm{harmonic}_{\mathrm{norm}}]}
{\mathbb{E}[\mathrm{harmonic}_{\mathrm{norm}}] + \mathbb{E}[\mathrm{eigenvector}]}, \\[10pt]
\beta_2=
\frac{\mathbb{E}[\mathrm{eigenvector}]}
{\mathbb{E}[\mathrm{harmonic}_{\mathrm{norm}}] + \mathbb{E}[\mathrm{eigenvector}]}.
\end{aligned}
\end{equation} 
\end{comment}

$$
\beta_1=
\frac{\mathbb{E}[\mathrm{eigenvector}]}
{\mathbb{E}[\mathrm{harmonic}_{\mathrm{norm}}] + \mathbb{E}[\mathrm{eigenvector}]},
\beta_2=
\frac{\mathbb{E}[\mathrm{harmonic}_{\mathrm{norm}}]}
{\mathbb{E}[\mathrm{harmonic}_{\mathrm{norm}}] + \mathbb{E}[\mathrm{eigenvector}]}
$$

This network-adaptive formulation of the hyperparameters requires no manual tuning and promotes balanced feature integration across networks with different sizes and topologies.

\section{Experimental protocol setup}
\label{sec:experimental}

%\subsection{Problem Formulation and Datasets}

\vspace{0.15cm}
\noindent\textit{Problem Formulation and Datasets.} We formulate influential node identification as a regression task, using continuous influence scores generated by the SIR model. Unlike binary classification, this formulation provides a more fine-grained representation of node influence. For each node $u$, we perform $T=1{,}000$\footnote{$T=1{,}000$ simulations are widely adopted in the literature for influential node identification.} independent SIR simulations, and define its ground-truth influence score as the average number of infected and recovered nodes at the end of the diffusion process. The infection probability is defined relative to the epidemic threshold as $\beta=\alpha\beta_{\mathrm{th}}$, with $\alpha\in[1,1.9]$, where
% , according to mean-field theory~\cite{macdonald2012spreaders},

$$
\beta_{\mathrm{th}}=
\frac{\langle k\rangle}
{\langle k^2\rangle-\langle k\rangle}.
$$

with $k$ denoting node degree, $\langle k\rangle$ the average node degree, and $\langle k^2\rangle$ the average squared degree over the network. The recovery rate is fixed at $\gamma=1$ throughout the simulations.

\vspace{0.15cm}
Experiments are conducted on a set of artificial and real-world networks spanning four domains: social, biological, technological, and knowledge. The datasets exhibit substantial variation in size, density, clustering, and degree distribution, enabling evaluation across heterogeneous network structures. Their main topological characteristics are reported in \textit{Table~\ref{tab:networks}}.% (with additional details provided in Table~\ref{tab:networks_} of the Appendix).

\begin{table*}[!h]
\centering
\caption{\textit{Topological properties of the network datasets.}}
\label{tab:networks}
\renewcommand{\arraystretch}{1.2}
\setlength{\tabcolsep}{2.5pt}
\scriptsize

\begin{tabular}{lcccc@{\hspace{10pt}}|lcccc}
\hline
\textbf{Network} & \textbf{Nodes} & \textbf{Edges} & \textbf{Avg. Deg.} & \textbf{$\Delta$} &
\textbf{Network} & \textbf{Nodes} & \textbf{Edges} & \textbf{Avg. Deg.} & \textbf{$\Delta$} \\
\hline

karate~\cite{kunegis2013konectsite} 
& 34 & 78 & 4.59 & 17 &
E\_road~\cite{vsubelj2011robust} 
& 1174 & 1417 & 2.41 & 10 \\

dolphins~\cite{kunegis2013konectsite} 
& 62 & 159 & 5.13 & 12 &
Faa~\cite{kunegis2013konect} 
& 1226 & 2408 & 3.93 & 34 \\

American\_football~\cite{kunegis2013konectsite} 
& 115 & 613 & 10.66 & 12 &
AS\_733~\cite{snapnets} 
& 6474 & 12572 & 3.88 & 1458 \\

Hep~\cite{leskovec2007graph} 
& 9877 & 25973 & 5.26 & 65 &
Figeys~\cite{ewing2007large} 
& 2239 & 6432 & 5.75 & 314 \\

GrQ~\cite{leskovec2007graph} 
& 5242 & 14484 & 5.53 & 81 &
Vidal~\cite{rual2005towards} 
& 3133 & 6149 & 3.93 & 129 \\

Netscience~\cite{newman2006finding} 
& 1461 & 2742 & 3.75 & 34 &
Bio\_Dmela~\cite{singh2008-isorank-multi} 
& 7393 & 25569 & 6.92 & 190 \\

Stelzl~\cite{stelzl2005human} 
& 1706 & 3155 & 3.70 & 95 &
David~\cite{meghanathan2015spectralradius} 
& 112 & 425 & 7.59 & 49 \\

soc\_anybeat~\cite{fire2012link} 
& 12645 & 49132 & 7.77 & 4800 &
soc\_slashdot~\cite{leskovec2009community} 
& 70068 & 358647 & 10.24 & 2507 \\

fb\_pages\_company~\cite{rozemberczki2019gemsec} 
& 14113 & 52126 & 7.39 & 215 &
PubMed~\cite{lee2023pgb} 
& 19717 & 44324 & 4.50 & 171 \\

web\_EPA~\cite{de2011exploratory} 
& 4271 & 8909 & 4.17 & 175 &
maayan\_pdzbase~\cite{beuming2005pdzbase} 
& 212 & 242 & 2.28 & 21 \\

web\_edu~\cite{gleich2004fast} 
& 3031 & 6474 & 4.27 & 104 &
Elegans~\cite{duch2005community} 
& 453 & 2025 & 8.94 & 237 \\

PowerGrid~\cite{dj1998collective} 
& 4941 & 6594 & 2.67 & 19 &
tvshow~\cite{rozemberczki2019gemsec} 
& 3892 & 17239 & 8.86 & 126 \\

\hline
\end{tabular}
\end{table*}

%\subsection{Evaluation Metrics and Baselines}

\vspace{0.15cm}
\noindent\textit{Evaluation Metrics and Baselines.} The ranking performance is evaluated using Kendall’s tau ($\tau$) \cite{kendall1938new}, Spearman’s rho ($\rho$) \cite{sedgwick2014spearman}, and NDCG \cite{jarvelin2002cumulated}, which respectively measure pairwise ranking consistency, monotonic agreement, and the accuracy of top-ranked influential nodes. The proposed framework \textit{CSBA-Net} is compared with three groups of baselines:

\begin{itemize}[noitemsep]
   \renewcommand{\labelitemi}{-} 
    \item \textit{Centrality-based methods}, including Degree Centrality (DC) \cite{6244633}, Betweenness Centrality (BC) \cite{brandes2001faster}, k-shell Centrality (KS) \cite{kitsak2010identification}, and H-index Centrality (HI) \cite{lu2016h}. 
    
    \item \textit{CNN-based models}, including RCNN \cite{YU2020105893}, M-RCNN \cite{OU2022117515}, and LCNN \cite{ahmad2024learning}; 
    
    \item \textit{Graph learning methods}, including InfGCN \cite{zhao2020infgcn} and GMHAR \cite{kou2023identify}. 
    \end{itemize}

All learning-based baselines are trained using the same SIR-derived ground-truth labels to ensure a fair comparison.

\vspace{0.15cm}
\noindent\textit{Implementation Details.} The proposed framework is implemented in PyTorch and evaluated on a workstation equipped with an NVIDIA GeForce RTX 2070 Max-Q (8 GB VRAM), Intel Core i7-10875H CPU (2.30 GHz), and 16 GB RAM.

The model uses a neighborhood size of $L=40$, a BiGRU hidden size of 128, and 4 attention heads. Training employs the Adam optimizer with a learning rate of $lr = 5\times10^{-4}$, batch size of 32, and 100 epochs ($N_{\mathrm{ep}} = 100$), using mini-batch training for efficient GPU utilization.

For a fair and consistent comparison, we follow the experimental protocol of Yu \textit{et al.} \cite{YU2020105893}, also adopted by M-RCNN \cite{OU2022117515} and LCNN \cite{ahmad2024learning}. The model is trained on a Barabási--Albert (BA) network BA(1000, 4), with $\langle k\rangle=4$. Yu \textit{et al.} \cite{YU2020105893} showed, on synthetic BA test networks, that training on a low average degree yields stable generalization across test networks of different densities, whereas training on higher average degrees degrades performance on sparser test networks. The trained model is then evaluated on diverse real-world networks without retraining.

% For a fair and consistent comparison, we follow the experimental protocol of Yu \textit{et al.} \cite{YU2020105893}, also adopted by M-RCNN \cite{OU2022117515} and LCNN~\cite{ahmad2024learning}. The model is trained on a Barabási--Albert (BA) network BA$(1000,4)$, with $\langle k\rangle=4$, a setting empirically demonstrated to generalize well across networks of different densities \cite{ahmad2024learning}. The trained model is then evaluated on diverse real-world networks without retraining.

% The proposed model is implemented using the following configuration. The Neighborhood size is $L = 40$, the BiGRU hidden size is $128$, and the number of attention heads is set to $4$. For the optimizer, we use Adam, and the learning rate is $5 \times 10^{-4}$, while the batch size is set to $32$. All hyperparameters are selected to ensure a balance between model capacity and generalization.

% The model is trained for a fixed number of epochs with early stopping based on validation loss. Also, to ensure reproducibility, all experiments are conducted using identical training settings across datasets and repeated multiple times with different random seeds.

\section{Results and Discussion}
\label{sec:results}

This section presents a comprehensive evaluation of the proposed framework across 24 benchmark networks from social, biological, technological, and knowledge domains. %Performance is assessed against centrality-based measures, CNN-based models, and graph neural network baselines using three complementary ranking metrics, namely Kendall's $\tau$, Spearman's $\rho$, and NDCG. 
We first compare the proposed approach with different classes of baselines to evaluate its ranking effectiveness and predictive capacity. Then, we conduct an ablation study to quantify the contribution of each architectural component, followed by a sensitivity analysis to examine the influence of diffusion intensity. Finally, a robustness analysis is performed to evaluate the stability and generalization of the proposed framework across diverse network categories.

\subsection{Overall Performance} \label{sec:overall}

We evaluate the ranking performance of CSBA-Net against the considered CNN-based, GNN-based, and centrality-based baselines.   Table~\ref{tab:overall_summary} reports the average Kendall’s $\tau$, Spearman’s $\rho$, and NDCG scores over the 24 benchmark networks and diffusion intensities $\alpha \in [1.0,1.9]$. CSBA-Net achieves the highest average value for all reported metrics, with $\bar{\tau}=0.802$, $\bar{\rho}=0.943$, NDCG@10 $=0.898$, and NDCG@20 $=0.916$. %ranking first in Kendall's $\tau$ on 23 out of 24 datasets and first in Spearman's $\rho$ on all 24 datasets.

 Among the CNN-based approaches, M-RCNN obtains the highest average Kendall’s $\tau$ of 0.677, followed by LCNN (0.644) and RCNN (0.600), \textit{i.e.}, CSBA-Net improves the average $\tau$ by approximately 18.5\%, 24.5\%, and 33.7\%, respectively. The average performance is also higher than that of the GNN-based methods, with $\tau=0.591$ for InfGCN and $\tau=0.538$ for GMHAR. Among the centrality baselines, H-index performs best ($\tau = 0.571$) yet still trails the proposed model, while betweenness centrality remains the weakest.% ($\tau = 0.431$).

\begin{table}[h!]
\centering
\caption{\textit{Average ranking scores across 24 datasets and $\alpha\in[1,1.9]$. Best values are shown in bold; NDCG is omitted for InfGCN and GMHAR due to their inductive evaluation protocol.
}}
\label{tab:overall_summary}
\footnotesize
\setlength{\tabcolsep}{5pt}
\renewcommand{\arraystretch}{1.1}
\begin{tabular}{l|lcccc}
\hline
\textbf{Class} & \textbf{Method} & \textbf{Kendall $\tau$} & \textbf{Spearman $\rho$} & \textbf{NDCG@10} & \textbf{NDCG@20} \\
\hline

\multirow{4}{*}{\textit{CNN-based}}
& RCNN & 0.600 & 0.761 & 0.806 & 0.811 \\
& M-RCNN & 0.677 & 0.838 & 0.848 & 0.846 \\
& LCNN & 0.644 & 0.778 & 0.841 & 0.838 \\
& CSBA-Net & \textbf{0.802} & \textbf{0.943} & \textbf{0.898} & \textbf{0.916} \\
\hline
\multirow{2}{*}{\textit{GNN-based}}
& InfGCN & 0.591 & 0.713 & -- & -- \\
& GMHAR & 0.538 & 0.692 & -- & -- \\
\hline
\multirow{4}{*}{\textit{Centrality}}
& DC & 0.560 & 0.731 & 0.743 & 0.767 \\
& BC & 0.431 & 0.592 & 0.628 & 0.669 \\
& HI & 0.571 & 0.740 & 0.764 & 0.778 \\
& KS & 0.541 & 0.710 & 0.728 & 0.753 \\

\hline
\end{tabular}
\end{table}

\begin{table*}[!h]
\centering
\caption{Average Kendall's $\tau$ per dataset, averaged over $\alpha \in [1.0,1.9]$.}
\label{tab:kendall_per_dataset}

\footnotesize
\setlength{\tabcolsep}{2pt}
\renewcommand{\arraystretch}{0.9}

\begin{tabular}{lcccccccccc}
\hline

\multirow{2}{*}{\textbf{Network}} &
\multicolumn{4}{c}{\textbf{CNN-based}} &
\multicolumn{2}{c}{\textbf{GNN-based}} &
\multicolumn{4}{c}{\textbf{Centrality}} \\

\cline{2-5}
\cline{6-7}
\cline{8-11}

& \textbf{CSBA-Net} & \textbf{RCNN} & \textbf{M-RCNN} & \textbf{LCNN}
& \textbf{InfGCN} & \textbf{GMHAR}
& \textbf{DC} & \textbf{BC} & \textbf{HI} & \textbf{KS} \\

\hline

\multicolumn{11}{l}{\textit{Social networks}} \\[1pt]
Karate & \textbf{0.912} & 0.661 & 0.769 & 0.798 & 0.624 & 0.774 & 0.656 & 0.568 & 0.602 & 0.533 \\
Dolphins & \textbf{0.938} & 0.631 & 0.770 & 0.864 & 0.610 & 0.466 & 0.701 & 0.517 & 0.768 & 0.510 \\
Am.\ Football & \textbf{0.677} & 0.417 & 0.502 & 0.150 & 0.486 & 0.380 & 0.532 & 0.210 & 0.418 & 0.105 \\
Anybeat & \textbf{0.686} & 0.571 & 0.644 & 0.198 & 0.617 & 0.390 & 0.415 & 0.332 & 0.416 & 0.417 \\
FB-Company & \textbf{0.829} & 0.747 & 0.777 & 0.695 & 0.705 & 0.636 & 0.544 & 0.417 & 0.567 & 0.566 \\
Slashdot & \textbf{0.779} & 0.555 & 0.658 & 0.283 & 0.564 & 0.456 & 0.670 & 0.603 & 0.682 & 0.683 \\
TVShow & 0.712 & \textbf{0.731} & 0.726 & 0.564 & 0.638 & 0.589 & 0.589 & 0.327 & 0.602 & 0.595 \\

\hline

\multicolumn{11}{l}{\textit{Biological networks}} \\[1pt]
Stelzl & \textbf{0.779} & 0.662 & 0.701 & 0.724 & 0.628 & 0.485 & 0.463 & 0.407 & 0.490 & 0.491 \\
Figeys & \textbf{0.685} & 0.589 & 0.663 & 0.500 & 0.567 & 0.487 & 0.477 & 0.440 & 0.495 & 0.498 \\
Vidal & \textbf{0.827} & 0.714 & 0.760 & 0.786 & 0.709 & 0.573 & 0.600 & 0.539 & 0.637 & 0.645 \\
Bio\_Dmela & \textbf{0.896} & 0.801 & 0.832 & 0.716 & 0.800 & 0.629 & 0.717 & 0.628 & 0.747 & 0.749 \\
PDZBase & \textbf{0.875} & 0.353 & 0.704 & 0.844 & 0.530 & 0.606 & 0.573 & 0.587 & 0.618 & 0.618 \\
Elegans & \textbf{0.879} & 0.835 & 0.860 & 0.812 & 0.568 & 0.457 & 0.649 & 0.483 & 0.660 & 0.667 \\

\hline

\multicolumn{11}{l}{\textit{Technological networks}} \\[1pt]
Web-EPA & \textbf{0.812} & 0.717 & 0.644 & 0.768 & 0.649 & 0.566 & 0.492 & 0.460 & 0.496 & 0.494 \\
Web\_Edu & \textbf{0.604} & 0.370 & 0.317 & 0.375 & 0.361 & 0.320 & 0.286 & 0.269 & 0.243 & 0.268 \\
PowerGrid & \textbf{0.653} & 0.365 & 0.499 & 0.575 & 0.530 & 0.550 & 0.354 & 0.243 & 0.372 & 0.328 \\
E\_Road & \textbf{0.858} & 0.056 & 0.338 & 0.649 & 0.562 & 0.731 & 0.461 & 0.313 & 0.480 & 0.486 \\
FAA & \textbf{0.847} & 0.650 & 0.722 & 0.789 & 0.643 & 0.582 & 0.500 & 0.263 & 0.564 & 0.571 \\
AS\_733 & \textbf{0.766} & 0.591 & 0.637 & 0.569 & 0.522 & 0.491 & 0.589 & 0.416 & 0.610 & 0.612 \\

\hline

\multicolumn{11}{l}{\textit{Knowledge networks}} \\[1pt]
HEP & \textbf{0.858} & 0.718 & 0.784 & 0.795 & 0.727 & 0.621 & 0.589 & 0.518 & 0.621 & 0.618 \\
GrQc & \textbf{0.697} & 0.650 & 0.672 & 0.630 & 0.617 & 0.503 & 0.627 & 0.423 & 0.633 & 0.624 \\
Netscience & \textbf{0.911} & 0.511 & 0.647 & 0.837 & 0.520 & 0.480 & 0.651 & 0.351 & 0.639 & 0.630 \\
PubMed & \textbf{0.788} & 0.672 & 0.696 & 0.636 & 0.514 & 0.648 & 0.445 & 0.371 & 0.478 & 0.489 \\
David & \textbf{0.974} & 0.827 & 0.920 & 0.888 & 0.482 & 0.500 & 0.853 & 0.649 & 0.871 & 0.793 \\

\hline

\textbf{Average} & \textbf{0.802} & 0.600 & 0.677 & 0.644 & 0.591 & 0.538 & 0.560 & 0.431 & 0.571 & 0.541 \\

\hline
\end{tabular}
\end{table*}

% TABLE  –  Spearman's rho, one row per network, all methods

\begin{table}[!h]
\centering
\caption{Average Spearman's $\rho$ per dataset, averaged over $\alpha \in [1.0,1.9]$.}
\label{tab:spearman_rho}
\footnotesize
\setlength{\tabcolsep}{2pt}
\renewcommand{\arraystretch}{0.9}

\begin{tabular}{lcccccccccc}
\hline

\multirow{2}{*}{\textbf{Network}} &
\multicolumn{4}{c}{\textbf{CNN-based}} &
\multicolumn{2}{c}{\textbf{GNN-based}} &
\multicolumn{4}{c}{\textbf{Centrality}} \\

\cline{2-5}
\cline{6-7}
\cline{8-11}

& \textbf{CSBA-Net} & \textbf{RCNN} & \textbf{M-RCNN} & \textbf{LCNN}
& \textbf{InfGCN} & \textbf{GMHAR}
& \textbf{DC} & \textbf{BC} & \textbf{HI} & \textbf{KS} \\

\hline

\multicolumn{11}{l}{\textit{Social networks}} \\[1pt]
Karate & \textbf{0.986} & 0.825 & 0.890 & 0.926 & 0.713 & 0.883 & 0.821 & 0.755 & 0.783 & 0.740 \\
Dolphins & \textbf{0.991} & 0.842 & 0.940 & 0.968 & 0.580 & 0.665 & 0.863 & 0.691 & 0.921 & 0.686 \\
Am.\ Football & \textbf{0.866} & 0.630 & 0.712 & 0.220 & 0.340 & 0.542 & 0.712 & 0.315 & 0.574 & 0.148 \\
Anybeat & \textbf{0.877} & 0.766 & 0.837 & 0.299 & 0.718 & 0.525 & 0.569 & 0.470 & 0.570 & 0.571 \\
FB\_Company & \textbf{0.973} & 0.909 & 0.926 & 0.842 & 0.848 & 0.822 & 0.730 & 0.587 & 0.750 & 0.749 \\
Slashdot & \textbf{0.952} & 0.741 & 0.832 & 0.444 & 0.730 & 0.593 & 0.849 & 0.782 & 0.858 & 0.859 \\
TVShow & 0.927 & \textbf{0.902} & 0.888 & 0.660 & 0.825 & 0.785 & 0.762 & 0.473 & 0.776 & 0.773 \\

\hline

\multicolumn{11}{l}{\textit{Biological networks}} \\[1pt]
Stelzl & \textbf{0.943} & 0.841 & 0.871 & 0.889 & 0.837 & 0.650 & 0.640 & 0.581 & 0.664 & 0.664 \\
Figeys & \textbf{0.880} & 0.785 & 0.848 & 0.669 & 0.788 & 0.658 & 0.660 & 0.620 & 0.675 & 0.678 \\
Vidal & \textbf{0.977} & 0.885 & 0.921 & 0.934 & 0.778 & 0.762 & 0.784 & 0.726 & 0.819 & 0.826 \\
Bio\_Dmela & \textbf{0.989} & 0.942 & 0.956 & 0.872 & 0.846 & 0.803 & 0.889 & 0.807 & 0.908 & 0.910 \\
PDZBase & \textbf{0.995} & 0.466 & 0.881 & 0.932 & 0.553 & 0.760 & 0.762 & 0.782 & 0.804 & 0.805 \\
Elegans & \textbf{0.990} & 0.962 & 0.969 & 0.946 & 0.577 & 0.601 & 0.830 & 0.652 & 0.841 & 0.852 \\

\hline

\multicolumn{11}{l}{\textit{Technological networks}} \\[1pt]
Web\_EPA & \textbf{0.967} & 0.878 & 0.816 & 0.918 & 0.837 & 0.738 & 0.675 & 0.641 & 0.678 & 0.678 \\
Web\_Edu & \textbf{0.763} & 0.508 & 0.420 & 0.485 & 0.504 & 0.450 & 0.393 & 0.369 & 0.333 & 0.369 \\
PowerGrid & \textbf{0.835} & 0.519 & 0.688 & 0.756 & 0.693 & 0.720 & 0.507 & 0.351 & 0.532 & 0.477 \\
E-Road & \textbf{0.990} & 0.058 & 0.496 & 0.827 & 0.731 & 0.798 & 0.630 & 0.450 & 0.656 & 0.675 \\
FAA & \textbf{0.989} & 0.850 & 0.897 & 0.938 & 0.833 & 0.719 & 0.683 & 0.386 & 0.748 & 0.757 \\
AS\_733 & \textbf{0.947} & 0.786 & 0.838 & 0.752 & 0.724 & 0.658 & 0.780 & 0.587 & 0.800 & 0.802 \\

\hline

\multicolumn{11}{l}{\textit{Knowledge networks}} \\[1pt]
HEP & \textbf{0.984} & 0.892 & 0.934 & 0.935 & 0.860 & 0.775 & 0.774 & 0.708 & 0.801 & 0.798 \\
GrQc & \textbf{0.882} & 0.837 & 0.851 & 0.738 & 0.733 & 0.660 & 0.804 & 0.597 & 0.812 & 0.805 \\
Netscience & \textbf{0.989} & 0.623 & 0.830 & 0.951 & 0.702 & 0.635 & 0.827 & 0.505 & 0.821 & 0.814 \\
PubMed & \textbf{0.953} & 0.860 & 0.870 & 0.789 & 0.686 & 0.733 & 0.621 & 0.536 & 0.656 & 0.665 \\
David & \textbf{0.993} & 0.962 & 0.990 & 0.979 & 0.669 & 0.665 & 0.969 & 0.825 & 0.978 & 0.943 \\

\hline

\textbf{Average} & \textbf{0.943} & 0.761 & 0.838 & 0.778 & 0.713 & 0.692 & 0.731 & 0.592 & 0.740 & 0.710 \\

\hline

\end{tabular}
\end{table}

\begin{figure*}[t]
\centering 
\begin{subfigure}{0.49\textwidth}
\centering
\includegraphics[width=\linewidth]{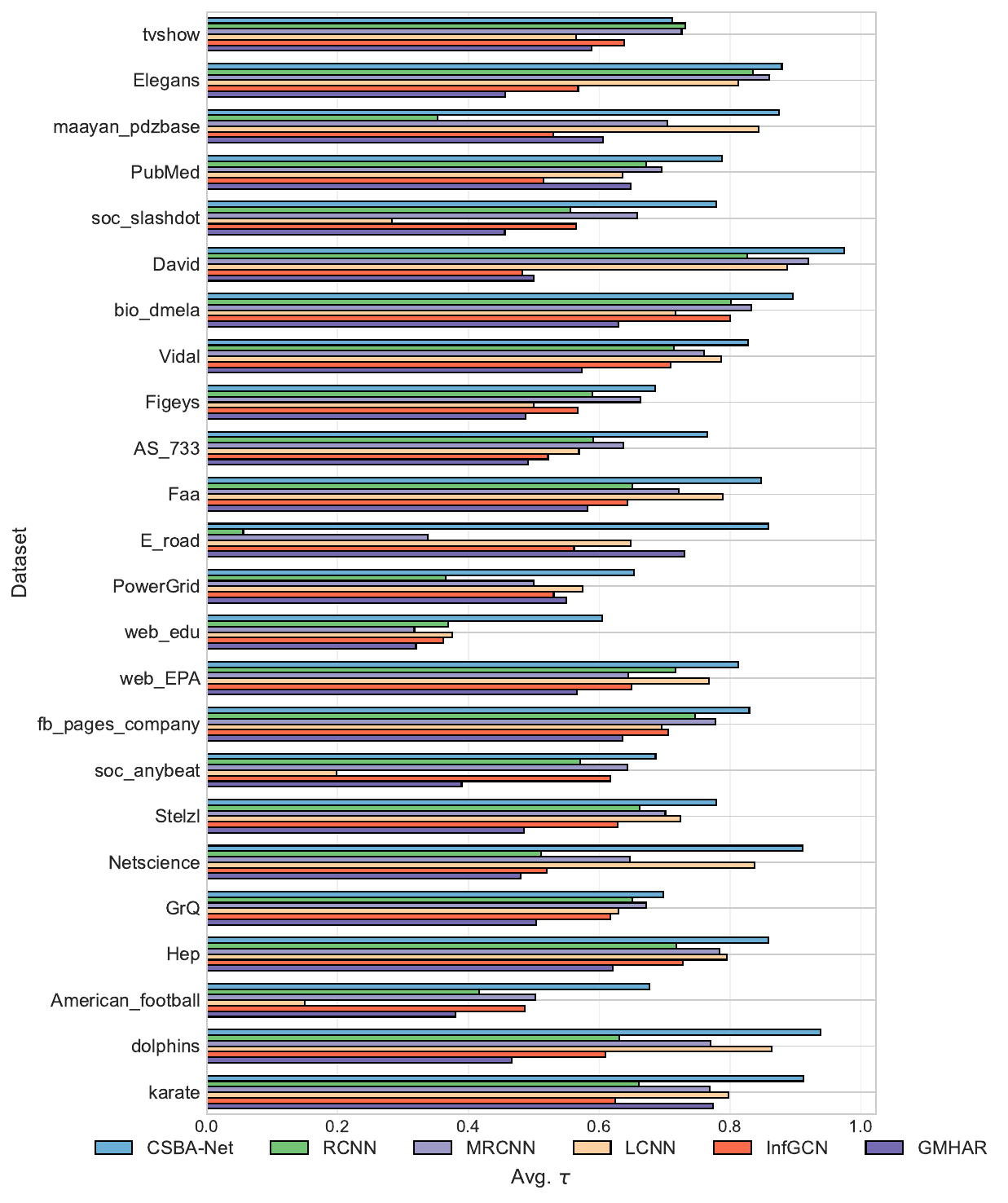}
\caption{Average Kendall $\tau$}
\label{fig:chart_av_tau}
\end{subfigure}
\hfill
\begin{subfigure}{0.49\textwidth}
\centering
\includegraphics[width=\linewidth]{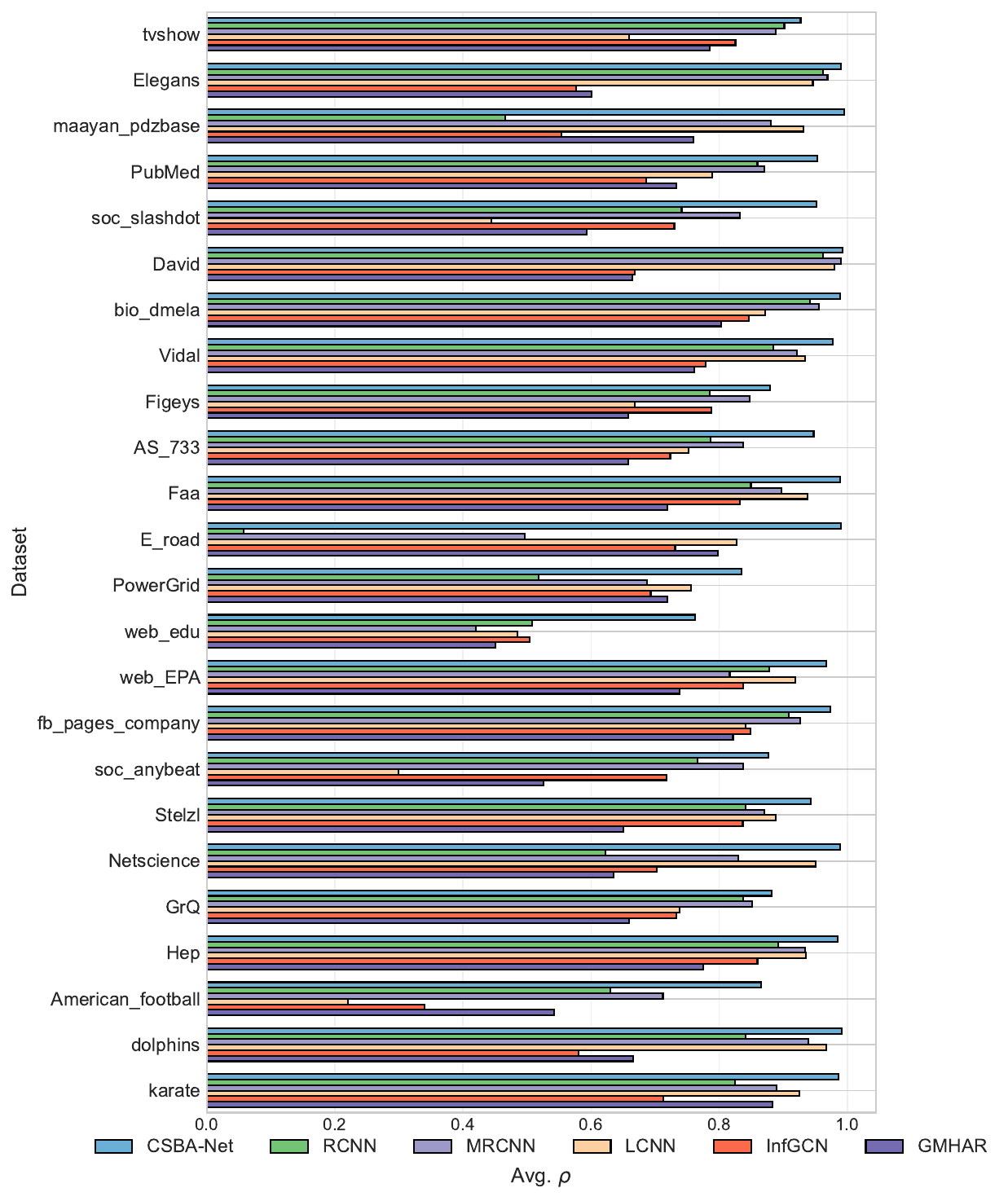} 
\caption{Average of Spearman $\rho$}
\label{fig:chart_av_Spearman}
\end{subfigure}
\caption{\textit{Performance comparison across datasets.}}
\label{fig:two_images_heatmaps_tau_rho}
\end{figure*}

\begin{figure*}[t]
\centering
\begin{subfigure}{0.49\textwidth}
\centering
\includegraphics[width=\linewidth]{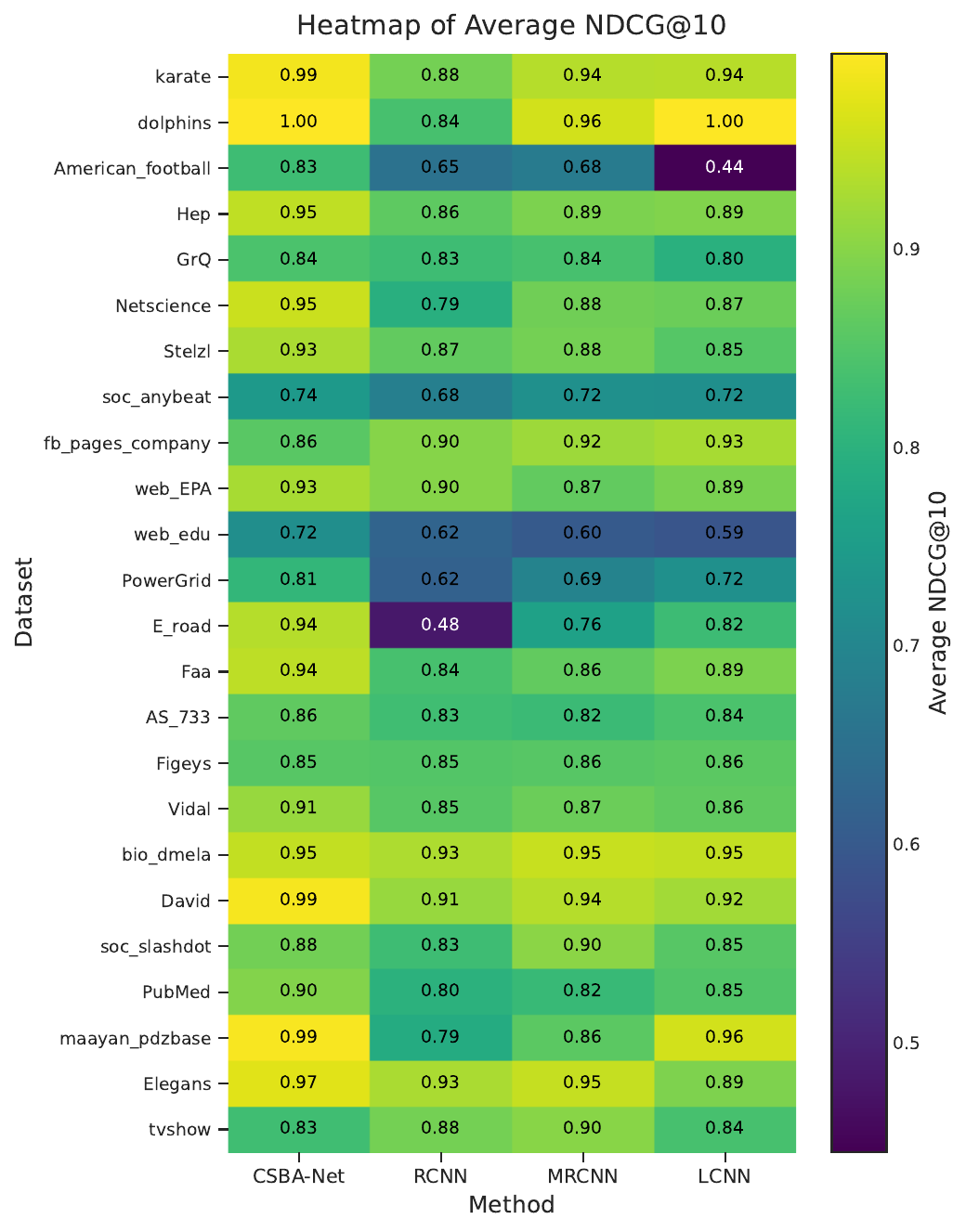}
\caption{Average NDCG@10}
\label{fig:ndcg10}
\end{subfigure}
\hfill
\begin{subfigure}{0.49\textwidth}
\centering
\includegraphics[width=\linewidth]{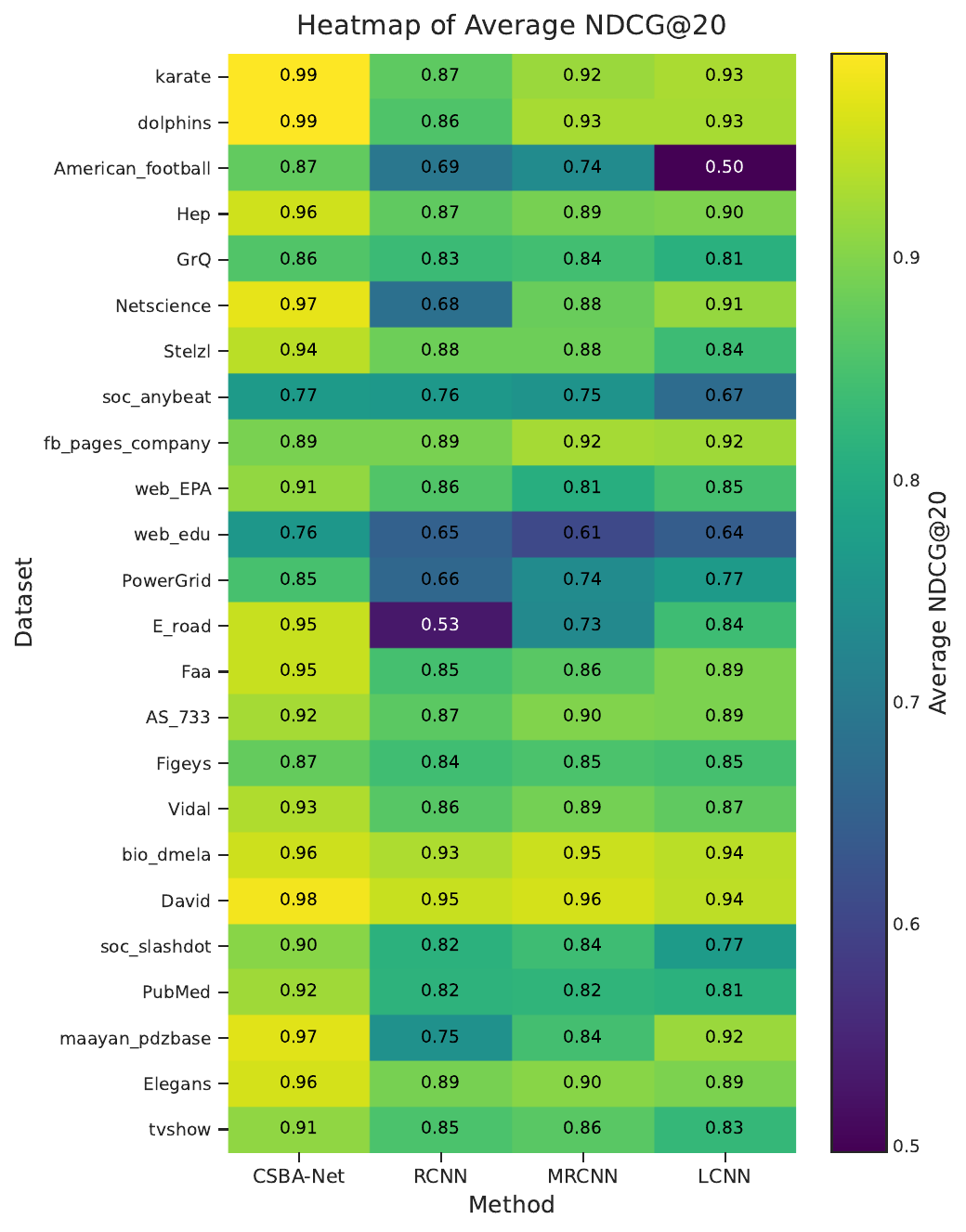}
\caption{Average NDCG@20}
\label{fig:ndcg20}
\end{subfigure}
\caption{Heatmap of average NDCG@10 (left) and NDCG@20 (right) per dataset, averaged over $\alpha\in[1.0,1.9]$. Higher values (brighter cells) indicate better top-$k$ ranking quality. CSBA-Net achieves the highest scores across the majority of datasets.}
\label{fig:ndcg}
\end{figure*}

Overall, the high values of both $\tau$ and $\rho$ indicate that the ordering predicted by CSBA-Net is well aligned with the reference ranking, \textit{i.e.} a strong ranking agreement with the SIR-derived ground-truth scores. Similarly, the NDCG (Table \ref{tab:overall_summary} and Figure \ref{fig:ndcg}) indicates that the model preserves the most influential nodes near the top of the ranking.

\subsection{Performance Across Network Structures and Domains%Analysis by Network Structure
} 
\label{sec:comparison}

In this section, we analyze how CSBA-Net performance varies with network structure, size, and sparsity.

\textit{Tables~\ref{tab:kendall_per_dataset}} and \textit{~\ref{tab:spearman_rho}} provide the per-network results and give a more detailed view of this aggregate performance. In general, CSBA-Net obtains the highest Kendall’s $\tau$ on 23 of the 24 benchmark networks and the highest Spearman’s $\rho$ on all 24 networks. It maintains high ranking correlations on both small networks, such as Karate and Dolphins, reaching $\rho = 0.986$ and $0.991$, and larger and more complex networks, such as Netscience and AS-733 ($\rho > 0.94$ in both cases). \textit{Figure~\ref{fig:two_images_heatmaps_tau_rho}} further illustrates the consistency of the ranking performance across the evaluated datasets.

% \label{sec:gnn_comparison}
% \begin{figure*}[t]
% \centering

% \begin{subfigure}{0.49\textwidth}
% \centering
% \includegraphics[width=\linewidth]{pictures/ndcg10_results_paper.pdf}
% \caption{Average NDCG@10}
% \label{fig:sub1}
% \end{subfigure}
% \hfill
% \begin{subfigure}{0.49\textwidth}
% \centering
% \includegraphics[width=\linewidth]{pictures/ndcg20_results_paper.pdf}
% % {pictures/heatmap_tau.pdf}
% \caption{Average NDCG@20}
% \label{fig:sub2}
% \end{subfigure} 

% \caption{Heatmap of average NDCG@10 (left) and NDCG@20 (right) per dataset, averaged over $\alpha \in [1.0, 1.9]$. Higher values (brighter cells) indicate better top-$k$ ranking quality. CSBA-Net achieves the highest scores across the majority of datasets.}  
% \label{fig:heatmap_ndcg}
% \end{figure*}

\vspace{0.15cm}
\noindent\textit{Performance on sparse and large-scale networks.} A larger performance difference is observed on sparse networks, where the average degree is relatively low and local neighborhoods contain fewer directly connected nodes. The results in \textit{Tables \ref{tab:kendall_per_dataset}} and \textit{\ref{tab:spearman_rho}} show that CSBA-Net maintains high ranking correlations on such graphs, whereas several CNN-based baselines show a larger degradation. This behavior is particularly visible on E-Road, PowerGrid, and Netscience. On E-Road, for example, CSBA-Net achieves a Kendall’s $\tau$ of 0.858, compared with 0.056 for RCNN (a relative gain exceeding $+1{,}400\%$), while on PowerGrid the corresponding values are 0.653 and 0.365 (+78.9\%). Netscience shows a similar pattern, with $\tau=0.911$ for CSBA-Net and 0.511 for RCNN.

Similarly, for large-scale networks, CSBA-Net maintains high ranking correlations. For instance, on \textit{soc\_slashdot}, with more than 70,000 nodes, it achieves $\tau = 0.779$ against $0.555$ for RCNN ($+40.4\%$) and $0.658$ for M-RCNN ($+18.4\%$), while LCNN deteriorates to $\tau = 0.283$.

%The model also generalises across network sizes, achieving mean $\tau=0.875$ on the five smallest networks (34–212 nodes), versus $0.733$ for M-RCNN and $0.578$ for RCNN. On the four largest networks (7{,}000–70{,}000 nodes), it maintains $\tau=0.830$, compared with $0.743$ and $0.686$, respectively, showing that its advantage persists as network size increases.

This behavior can be attributed to the complementary signals added by CSBA-Net. In fact, CSBA-Net incorporates meso- and macro-level structural descriptors, including community-bridging and global-reachability information, which provide useful information to identify influential nodes. This is noteworthy for sparse networks, where the immediate neighborhood of a node contains relatively few connections, and local connectivity patterns provide less information to distinguish nodes with different spreading potential. Similarly, for 
large-scale networks, such representations can retain information about a node's position in the broader network even when the number of nodes and possible diffusion paths increases substantially.

\vspace{0.15cm}
\noindent\textit{Scalability and structural variation on Barabási–Albert networks.} To examine the effect of network size and density under more controlled conditions, \textit{Tables~\ref{tab:ba_kendall}} and\textit{~\ref{tab:ba_spearman}} %and Fig. \ref{fig:knd_tau_ba} 
report the performance on Barabási–Albert (BA) networks ranging from $1{,}000$ to $6{,}000$ nodes with average degrees of $2$, $4$, $8$, and $10$. The sparse series with $\langle k\rangle=2$ show that as the network size increases from 1,000 to 6,000 nodes, Kendall’s $\tau$ for CSBA-Net decreases from 0.909 to 0.876, whereas RCNN decreases from 0.819 to 0.741 and M-RCNN from 0.820 to 0.722. A similar trend is observed for Spearman’s $\rho$, where CSBA-Net changes from 0.996 to 0.983, compared with a decrease from 0.956 to 0.904 for RCNN and from 0.957 to 0.888 for M-RCNN. Thus, CSBA-Net is not completely insensitive to increasing network size, but its ranking performance exhibits a smaller degradation than the CNN-based baselines. For the remaining BA configurations ($\langle k\rangle\in\{4,8,10\}$), the absolute correlation values vary across network sizes and densities for all methods, without a monotonic dependence on density, while CSBA-Net remains the best-performing method in every configuration.

% For denser BA configurations, we observe that the absolute correlation values decrease for all methods, indicating that increasing connectivity does not make influential-node ranking easier. 

On the other hand, for some BA networks, the performance of CSBA-Net depends on the scale-free level of the network. In fact, when a relatively small number of hubs coexist with many low-degree nodes, local structural descriptors can provide highly heterogeneous information across nodes. The multi-scale representations used by CSBA-Net provide additional global and community-level descriptors that may help distinguish nodes whose local connectivity patterns are similar. 

%The results are therefore consistent with the hypothesis that combining complementary structural scales is beneficial for ranking influential nodes in heterogeneous BA topologies.

Overall, the BA experiments complement the real-world evaluation, and show that CSBA-Net achieves the highest Kendall’s $\tau$ and Spearman’s $\rho$ across different configurations, and hence retains high ranking agreement as both network size and average degree vary.

\vspace{0.15cm}
\noindent\textit{Cross-domain generalization.} We further assess the performance of CSBA-Net across different network domains, which exhibit diverse structural and functional characteristics. For each domain, we compute the average Kendall’s $\tau$  across the corresponding networks to obtain an overall measure of ranking consistency.
CSBA-Net achieves mean Kendall’s $\tau$ values of $0.790$, $0.823$, $0.757$, and $0.846$ for social, biological, technological, and knowledge networks, respectively. It outperforms the best competing method in each category, namely M-RCNN on social and biological domains, and LCNN on technological and knowledge domains, with gains ranging from $9.3\%$ to $21.9\%$. The variation in average performance across domains highlights the impact of network characteristics on influential-node identification. In particular, the largest gap occurs in technological networks, suggesting they may present more challenging structural properties. RCNN achieves a mean $\tau=0.458$, versus $0.757$ for CSBA-Net (+65.2\%). On the challenging Web\_Edu network ($\langle k\rangle=4.27$), CSBA-Net also leads LCNN ($0.604$ vs.~$0.375$), a gain of $+61.1\%$.

These consistent gains over baselines support the ability of CSBA-Net's multi-scale structural representations to capture topology relevant to node influence across diverse network contexts.

%This also explains the sharp degradation of LCNN on large heterogeneous graphs which follows directly from its strict one-hop restriction, and prevents it from capturing influence that propagates beyond the immediate vicinity of a node. As network size and structural heterogeneity increase, this locality bias becomes increasingly penalizing.

%% ============================================================
%%  TABLE A — BA Networks: Kendall τ  (CNN + GNN + Δ)
%% ============================================================

\begin{table*}[t]
\centering
\caption{\textit{Average Kendall~$\tau$ on synthetic Barab\'{a}si--Albert (BA) networks
of increasing size and varying average degree.
$\Delta X = (\tau_{\text{CSBA-Net}} - \tau_{X})/\tau_{X} \times 100$.
Results averaged over $\alpha \in [1.0, 1.9]$. Best value per row in bold.}}
\label{tab:ba_kendall}

\scriptsize
\setlength{\tabcolsep}{2.5pt}
\renewcommand{\arraystretch}{0.80}
\resizebox{\textwidth}{!}{%

\begin{tabular}{l|cccccc|ccccc}
\hline
 & \multicolumn{6}{c|}{\textbf{Average Kendall~$\tau$}}
 & \multicolumn{5}{c}{\textbf{Relative gain of CSBA-Net (\%)}} \\
\cline{2-7}\cline{8-12}
\textbf{Network}
  & \textbf{CSBA-Net} & \textbf{RCNN} & \textbf{M-RCNN} & \textbf{LCNN}
  & \textbf{InfGCN} & \textbf{GMHAR}
  & $\Delta$\textbf{RCNN} & $\Delta$\textbf{M-RCNN} & $\Delta$\textbf{LCNN}
  & $\Delta$\textbf{InfGCN} & $\Delta$\textbf{GMHAR} \\
\hline
\multicolumn{12}{l}{\textit{1\,000 nodes}} \\
BA-1k-2   & \textbf{0.909} & 0.819 & 0.820 & 0.866 & 0.343 & 0.295
          & $+11.0\%$ & $+10.9\%$ & $+5.0\%$   & $+165.0\%$ & $+207.9\%$ \\
BA-1k-4   & \textbf{0.323} & 0.275 & 0.274 & 0.283 & 0.243 & 0.038
          & $+17.7\%$ & $+18.0\%$ & $+14.1\%$  & $+32.9\%$  & $+755.7\%$ \\
BA-1k-8   & \textbf{0.836} & 0.807 & 0.807 & 0.760 & 0.410 & 0.407
          & $+3.5\%$  & $+3.5\%$  & $+9.9\%$   & $+103.9\%$ & $+105.4\%$ \\
BA-1k-10  & \textbf{0.530} & 0.457 & 0.470 & 0.434 & 0.427 & 0.283
          & $+15.9\%$ & $+12.7\%$ & $+22.2\%$  & $+24.1\%$  & $+87.4\%$  \\
\hline
\multicolumn{12}{l}{\textit{2\,000 nodes}} \\
BA-2k-2   & \textbf{0.896} & 0.805 & 0.810 & 0.843 & 0.735 & 0.319
          & $+11.3\%$ & $+10.5\%$ & $+6.3\%$   & $+21.9\%$  & $+180.6\%$ \\
BA-2k-4   & \textbf{0.907} & 0.805 & 0.818 & 0.850 & 0.759 & 0.312
          & $+12.7\%$ & $+10.9\%$ & $+6.7\%$   & $+19.5\%$  & $+190.4\%$ \\
\hline
\multicolumn{12}{l}{\textit{3\,000 nodes}} \\
BA-3k-2   & \textbf{0.902} & 0.806 & 0.813 & 0.848 & 0.749 & 0.314
          & $+11.9\%$ & $+10.9\%$ & $+6.3\%$   & $+20.5\%$  & $+187.5\%$ \\
BA-3k-4   & \textbf{0.303} & 0.263 & 0.266 & 0.276 & 0.255 & 0.180
          & $+15.2\%$ & $+14.0\%$ & $+10.0\%$  & $+18.7\%$  & $+68.5\%$  \\
BA-3k-10  & \textbf{0.467} & 0.360 & 0.378 & 0.365 & 0.413 & 0.234
          & $+29.7\%$ & $+23.6\%$ & $+27.8\%$  & $+13.0\%$  & $+99.8\%$  \\
\hline
\multicolumn{12}{l}{\textit{4\,000--6\,000 nodes (sparse, $\langle k\rangle=2$)}} \\
BA-4k-2   & \textbf{0.901} & 0.799 & 0.810 & 0.843 & 0.662 & 0.356
          & $+12.8\%$ & $+11.3\%$ & $+6.9\%$   & $+36.1\%$  & $+153.5\%$ \\
BA-5k-2   & \textbf{0.890} & 0.772 & 0.762 & 0.838 & 0.673 & 0.325
          & $+15.2\%$ & $+16.8\%$ & $+6.2\%$   & $+32.2\%$  & $+173.7\%$ \\
BA-6k-2   & \textbf{0.876} & 0.741 & 0.722 & 0.823 & 0.673 & 0.305
          & $+18.2\%$ & $+21.3\%$ & $+6.4\%$   & $+30.0\%$  & $+187.6\%$ \\
\hline
\end{tabular}}
\end{table*}

%% ============================================================
%%  TABLE B — BA Networks: Spearman ρ  (CNN + GNN + Δ)
%% ============================================================
\begin{table*}[t]
\centering
\caption{\textit{Average Spearman~$\rho$ on synthetic Barab\'{a}si--Albert (BA) networks
of increasing size and varying average degree.
$\Delta X = (\rho_{\text{CSBA-Net}} - \rho_{X})/\rho_{X} \times 100$.
Results averaged over $\alpha \in [1.0, 1.9]$. Best value per row in bold.}}
\label{tab:ba_spearman}
\scriptsize
\setlength{\tabcolsep}{2.5pt} 
\resizebox{\textwidth}{!}{%

\begin{tabular}{l|cccccc|ccccc}
\hline
 & \multicolumn{6}{c|}{\textbf{Average Spearman~$\rho$}}
 & \multicolumn{5}{c}{\textbf{Relative gain of CSBA-Net (\%)}} \\
\cline{2-7}\cline{8-12}
\textbf{Network}
  & \textbf{CSBA-Net} & \textbf{RCNN} & \textbf{M-RCNN} & \textbf{LCNN}
  & \textbf{InfGCN} & \textbf{GMHAR}
  & $\Delta$\textbf{RCNN} & $\Delta$\textbf{M-RCNN} & $\Delta$\textbf{LCNN}
  & $\Delta$\textbf{InfGCN} & $\Delta$\textbf{GMHAR} \\
\hline
\multicolumn{12}{l}{\textit{1\,000 nodes}} \\
BA-1k-2   & \textbf{0.996} & 0.956 & 0.957 & 0.956 & 0.603 & 0.475
          & $+4.2\%$  & $+4.1\%$  & $+4.2\%$   & $+65.2\%$  & $+109.8\%$ \\
BA-1k-4   & \textbf{0.449} & 0.399 & 0.398 & 0.410 & 0.343 & 0.057
          & $+12.6\%$ & $+12.8\%$ & $+9.5\%$   & $+30.8\%$  & $+693.6\%$ \\
BA-1k-8   & \textbf{0.981} & 0.949 & 0.950 & 0.919 & 0.610 & 0.569
          & $+3.4\%$  & $+3.3\%$  & $+6.8\%$   & $+60.9\%$  & $+72.3\%$  \\
BA-1k-10  & \textbf{0.712} & 0.633 & 0.650 & 0.607 & 0.574 & 0.411
          & $+12.4\%$ & $+9.6\%$  & $+17.4\%$  & $+24.0\%$  & $+73.4\%$  \\
\hline
\multicolumn{12}{l}{\textit{2\,000 nodes}} \\
BA-2k-2   & \textbf{0.992} & 0.947 & 0.951 & 0.968 & 0.795 & 0.453
          & $+4.8\%$  & $+4.3\%$  & $+2.5\%$   & $+24.8\%$  & $+118.9\%$ \\
BA-2k-4   & \textbf{0.996} & 0.949 & 0.956 & 0.970 & 0.865 & 0.447
          & $+5.0\%$  & $+4.2\%$  & $+2.6\%$   & $+15.1\%$  & $+122.9\%$ \\
\hline
\multicolumn{12}{l}{\textit{3\,000 nodes}} \\
BA-3k-2   & \textbf{0.994} & 0.949 & 0.954 & 0.970 & 0.865 & 0.446
          & $+4.7\%$  & $+4.2\%$  & $+2.5\%$   & $+14.9\%$  & $+123.0\%$ \\
BA-3k-4   & \textbf{0.422} & 0.381 & 0.386 & 0.399 & 0.345 & 0.119
          & $+10.6\%$ & $+9.3\%$  & $+5.7\%$   & $+22.3\%$  & $+253.8\%$ \\
BA-3k-10  & \textbf{0.637} & 0.510 & 0.533 & 0.515 & 0.590 & 0.344
          & $+24.9\%$ & $+19.4\%$ & $+23.6\%$  & $+8.0\%$   & $+85.3\%$  \\
\hline
\multicolumn{12}{l}{\textit{4\,000--6\,000 nodes (sparse, $\langle k\rangle=2$)}} \\
BA-4k-2   & \textbf{0.994} & 0.945 & 0.951 & 0.946 & 0.891 & 0.505
          & $+5.1\%$  & $+4.4\%$  & $+5.1\%$   & $+11.5\%$  & $+96.6\%$  \\
BA-5k-2   & \textbf{0.989} & 0.927 & 0.920 & 0.945 & 0.905 & 0.464
          & $+6.8\%$  & $+7.6\%$  & $+4.7\%$   & $+9.3\%$   & $+113.2\%$ \\
BA-6k-2   & \textbf{0.983} & 0.904 & 0.888 & 0.939 & 0.910 & 0.433
          & $+8.8\%$  & $+10.8\%$ & $+4.8\%$   & $+8.1\%$   & $+127.2\%$ \\
\hline
\end{tabular}}
\end{table*}

\subsection{Ablation Study} 
 \label{sec:ablation}

To better understand the contribution of each architectural component to the overall performance of our model, we conduct a controlled ablation study by systematically removing or replacing key modules from the full architecture. Four ablated configurations are evaluated alongside the complete model:

\begin{enumerate}[label=(\roman*),noitemsep]
    \item \textit{CSBA-Net\_NoBiGRU}, in which the BiGRU-based cross-scale fusion is replaced by direct concatenation; 
    \item \textit{CSBA-Net\_NoAttention}, in which the multi-head self-attention layer is removed; 
    \item \textit{CSBA-Net\_SingleScale}, which uses only the micro-level representation $D^u$; and 
    \item \textit{CSBA-Net\_RandomNeighbor}, in which the proposed structured neighborhood selection is replaced by uniform random sampling. All other components and the evaluation protocol are kept unchanged.
\end{enumerate}

\begin{comment}
\begin{itemize}[noitemsep]
    \item[(i)] \texttt{CSBA-Net\_NoBiGRU}: the BiGRU sequential fusion module is removed and replaced by direct concatenation of the three CNN-derived embeddings, reducing the inter-scale sequence of dimension $(B, 3, 32)$ to a flat vector of dimension $(B, 96)$ fed directly to the fully connected layers. This configuration eliminates all explicit cross-scale dependency modeling.

    \item[(ii)] \texttt{CSBA-Net\_NoAttention}: the multi-head self-attention layer is removed while the BiGRU module is retained. The BiGRU output of dimension $(B, 3, 256)$ is mean-pooled directly over the sequence dimension and passed to the fully connected layers, suppressing adaptive scale reweighting.

    \item[(iii)] \texttt{CSBA-Net\_SingleScale}: the three-scale structural input is reduced to the micro-level matrix $D^u$ exclusively. The BiGRU and multi-head attention layers are retained but receive a degenerate sequence of length one, $(B, 1, 32)$, which collapses inter-scale modeling to a trivial operation.

    \item[(iv)] \texttt{CSBA-Net\_RandomNeighbor}: the proposed betweenness--harmonic BFS-based neighborhood selection is replaced by uniform random sampling, while the full three-scale architecture is otherwise preserved. This configuration isolates the contribution of the structured subgraph construction strategy.
\end{itemize}
\end{comment}

From \textit{Table~\ref{tab:ablation_kendall}}, removing the BiGRU module causes the largest performance degradation among the evaluated variants, with average Kendall’s $\tau$ decreasing from $0.8018$ to $0.4216$, corresponding to a relative drop of $47.4\%$. The degradation is particularly pronounced on several networks, where the ranking performance approaches low or even negative correlation with the SIR-based ground truth. As the BiGRU is used to capture the cross-scale dependencies, these results show the usefulness of explicitly modeling interactions between the three structural scales, and that a simple concatenation of the scale-specific features is insufficient to reproduce the performance of the full model.

The single-scale variant, which relies only on $D^u$, produces the second-largest degradation, with $\tau$ falling to $0.5878$ ($-26.7\%$). This drop is particularly pronounced on networks where global structural position carries substantial information about diffusion potential, such as \textit{GrQc}, \textit{PubMed}, and \textit{HEP}. This is because in these large, sparsely connected networks, micro-level connectivity patterns alone are less discriminative as many nodes share similar local degree and clustering profiles. On the other hand, these nodes differ substantially in their global reachability and bridging roles, properties that are only captured by $B^u$ and $H^u$.

Replacing the structured neighborhood selection with random sampling reduces the average $\tau$ to 0.6901 ($-13.9\%$), especially on networks with skewed degree distributions or strong community structure, such as \textit{HEP}, \textit{Elegans}, and \textit{GrQc}. This confirms that the betweenness–harmonic ranking provides more informative local contexts and preserves structural patterns that are more relevant to influence estimation, whereas random sampling weakens this signal and introduces less informative neighborhoods.

Finally, removing the multi-head self-attention layer leads to the smallest, though still consistent, decline, with $\tau$ decreasing to $0.7234$ ($-9.8\%$). The decline is relatively uniform across different networks. Without attention, the BiGRU output is passed forward, with each scale contributing according to a fixed, learned weighting rather than being adaptively reweighted according to the characteristics of each node's neighborhood. The smaller drop compared with removing the BiGRU indicates that cross-scale sequential modeling has a larger contribution to the overall ranking performance, while the attention mechanism provides an additional refinement of the learned representation.

%Similarly, the decrease observed without attention indicates that adaptive reweighting of the learned scale representations provides an additional benefit after cross-scale feature extraction and fusion.

%Lacking this fusion, plain concatenation leaves the FC layers without a mechanism to resolve which scale should dominate; attention only fine-tunes an already fused representation, and neighbor selection governs the quality of the raw input entering this pipeline.

%In summary, each of the four architectural elements contributes meaningfully and non-redundantly to overall performance, with the severity of BiGRU removal underscoring that explicit, ordered cross-scale fusion is the cornerstone of the proposed design.
% \begin{table*}[t]

\begin{table}[!htbp]

\centering
\caption{%Ablation study: average Kendall's $\tau$ per dataset, averaged over $\alpha \in [1.0, 1.9]$. Best value per row in bold. "w/o BiGRU" replaces the BiGRU with direct concatenation; "w/o Attention" removes the multi-head self-attention layer; "Single-Scale" uses only the micro-level matrix $D^{u}$; "Random Neighbor" replaces structural-based BFS selection with random sampling.
\textit{Ablation study: average Kendall’s $\tau$ per dataset over $\alpha \in [1.0, 1.9]$. Best values per row are in bold. \textit{w/o BiGRU} replaces the BiGRU with direct concatenation; \textit{w/o Attn} removes multi-head self-attention; \textit{Single-Scale} uses only the micro-level matrix $D^{u}$; \textit{Random Nbr} replaces BFS-based structural selection with random sampling.}}
\label{tab:ablation_kendall}
%\scriptsize
\footnotesize
\setlength{\tabcolsep}{5pt}
\begin{tabular}{lrrrrr} 
\hline
\textbf{Network}
  & \textbf{CSBA-Net (Full)}
  & \textbf{w/o Attn}
  & \textbf{w/o BiGRU}
  & \textbf{Single-Scale}
  & \textbf{Random Nbr} \\
\hline
\multicolumn{6}{l}{\textit{Social networks}} \\
  Karate          & \textbf{0.9120} & 0.8178 & 0.7655 & 0.7977 & 0.8092 \\
  Dolphins        & \textbf{0.9378} & 0.8465 & 0.8152 & 0.7669 & 0.8552 \\
  Am.\ Football   & \textbf{0.6773} & 0.6093 & 0.3817 & 0.5264 & 0.5895 \\
  Soc\_anybeat     & \textbf{0.6861} & 0.6158 & $-$0.2679 & 0.4985 & 0.5159 \\
  FB\_Company      & \textbf{0.8294} & 0.7488 & 0.4644 & 0.6064 & 0.7560 \\
  Soc\_slashdot    & \textbf{0.7786} & 0.7233 & $-$0.3522 & 0.6369 & 0.6031 \\
  TVShow          & \textbf{0.7115} & 0.6259 & 0.0471 & 0.3535 & 0.6531 \\
\hline
\multicolumn{6}{l}{\textit{Biological networks}} \\
  Stelzl          & \textbf{0.7786} & 0.7255 & 0.1921 & 0.4972 & 0.6967 \\
  Figeys          & \textbf{0.6850} & 0.6274 & 0.1610 & 0.5809 & 0.5711 \\
  Vidal           & \textbf{0.8271} & 0.7575 & 0.6745 & 0.6460 & 0.7343 \\
  Bio\_Dmela       & \textbf{0.8958} & 0.8321 & 0.3756 & 0.7783 & 0.8011 \\
  PDZBase         & \textbf{0.8747} & 0.7525 & 0.7302 & 0.7108 & 0.7251 \\
  Elegans         & \textbf{0.8792} & 0.8094 & 0.1827 & 0.5216 & 0.7504 \\
\hline
\multicolumn{6}{l}{\textit{Technological networks}} \\
  Web\_EPA         & \textbf{0.8122} & 0.7102 & 0.6687 & 0.6615 & 0.7204 \\
  Web\_Edu         & \textbf{0.6044} & 0.4519 & 0.2819 & 0.3333 & 0.4849 \\
  PowerGrid       & \textbf{0.6530} & 0.6076 & 0.5721 & 0.5441 & 0.6062 \\
  E\_Road          & \textbf{0.8582} & 0.7721 & 0.7053 & 0.6749 & 0.7146 \\
  FAA             & \textbf{0.8473} & 0.7655 & 0.7146 & 0.7119 & 0.7214 \\
  AS\_733          & \textbf{0.7655} & 0.6921 & $-$0.2660 & 0.5840 & 0.5476 \\
\hline
\multicolumn{6}{l}{\textit{Knowledge networks}} \\
  Hep             & \textbf{0.8583} & 0.7865 & 0.5742 & 0.4866 & 0.7160 \\
  GrQc            & \textbf{0.6974} & 0.6163 & 0.3504 & 0.2096 & 0.6215 \\
  Netscience      & \textbf{0.9107} & 0.8415 & 0.7578 & 0.7126 & 0.7637 \\
  PubMed          & \textbf{0.7879} & 0.7228 & 0.7488 & 0.4201 & 0.7310 \\
  David           & \textbf{0.9741} & 0.9037 & 0.8399 & 0.8483 & 0.8753 \\
\hline
\textbf{Average}  & \textbf{0.8018} & 0.7234 & 0.4216 & 0.5878 & 0.6901 \\
$\Delta$ (\%)     & ---             & $-$9.8 & $-$47.4 & $-$26.7 & $-$13.9 \\
\hline
\end{tabular}
\end{table}

All these results, together, support the complementary roles of multi-scale structural representations, structured neighborhood selection, sequential cross-scale fusion, and attention in the proposed architecture. In other words, $D^u$, $B^u$, and $H^u$ encode complementary but individually incomplete views of a node's structural context.

%The BiGRU's role is to fuse them into an ordered representation in which macro- and micro-level cues mutually condition one another. 

%The degradation observed in the SingleScale variant suggests that local connectivity and clustering information alone is insufficient to reproduce the performance of the full model, while the additional meso- and macro-level representations provide useful complementary signals. 

%The lower performance of the RandomNeighbor variant further supports the usefulness of the proposed structured neighborhood construction, although this experiment does not isolate the individual contributions of betweenness and harmonic centrality from the overall selection strategy. 

\begin{figure*}[!htbp] 
    \centering 
    \includegraphics[width=0.9\textwidth]{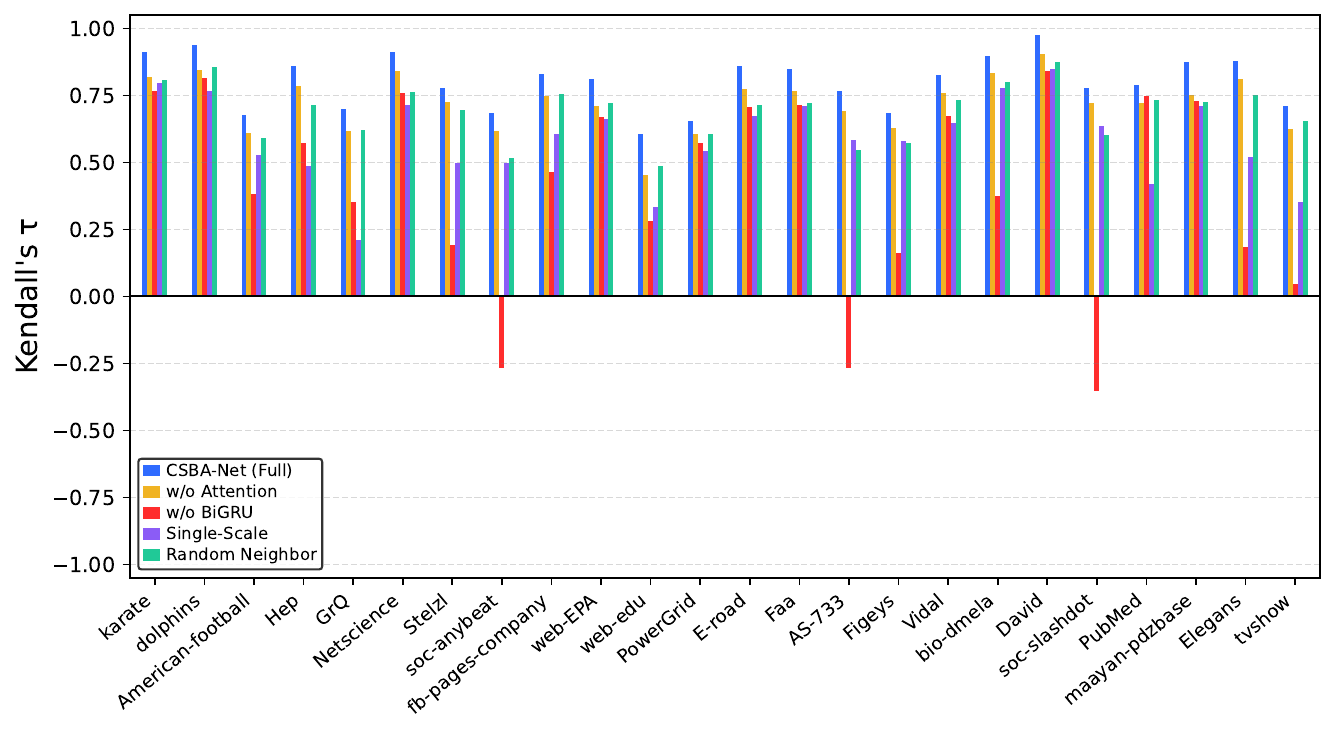}
\caption{\textit{Illustration of the average Kendall’s $\tau$ per dataset over $\alpha \in [1.0, 1.9]$ according to the Ablation study results in Table~\ref{tab:ablation_kendall}}. } 

    \label{fig:ablation_bar_ArgTau}
\end{figure*}

\textit{Figure~\ref{fig:ablation_bar_ArgTau}} illustrates these differences more clearly and shows that they are not confined to a single dataset.  

\subsection{Sensitivity and Stability to Diffusion Intensity%Sensitivity Analysis
}

Assessing model stability across diffusion conditions is essential for establishing its practical reliability. We evaluate sensitivity of CSBA-Net to the diffusion intensity by varying the SIR transmission parameter through the scaling factor $\alpha$. Specifically, the transmission probability is defined as $\beta=\alpha\beta_{\text{th}}$, where $\alpha$ controls the diffusion intensity and is varied from 1.0 to 1.9. We therefore analyze the evolution of Kendall’s $\tau$ as a function of $\alpha$. Note that, unlike threshold-based binary classification approaches, our regression framework produces continuous influence scores, requiring rankings to remain consistent with the ground truth across diffusion conditions.

\begin{figure*}[htbp]
    \centering
    \includegraphics[width=0.9\textwidth]{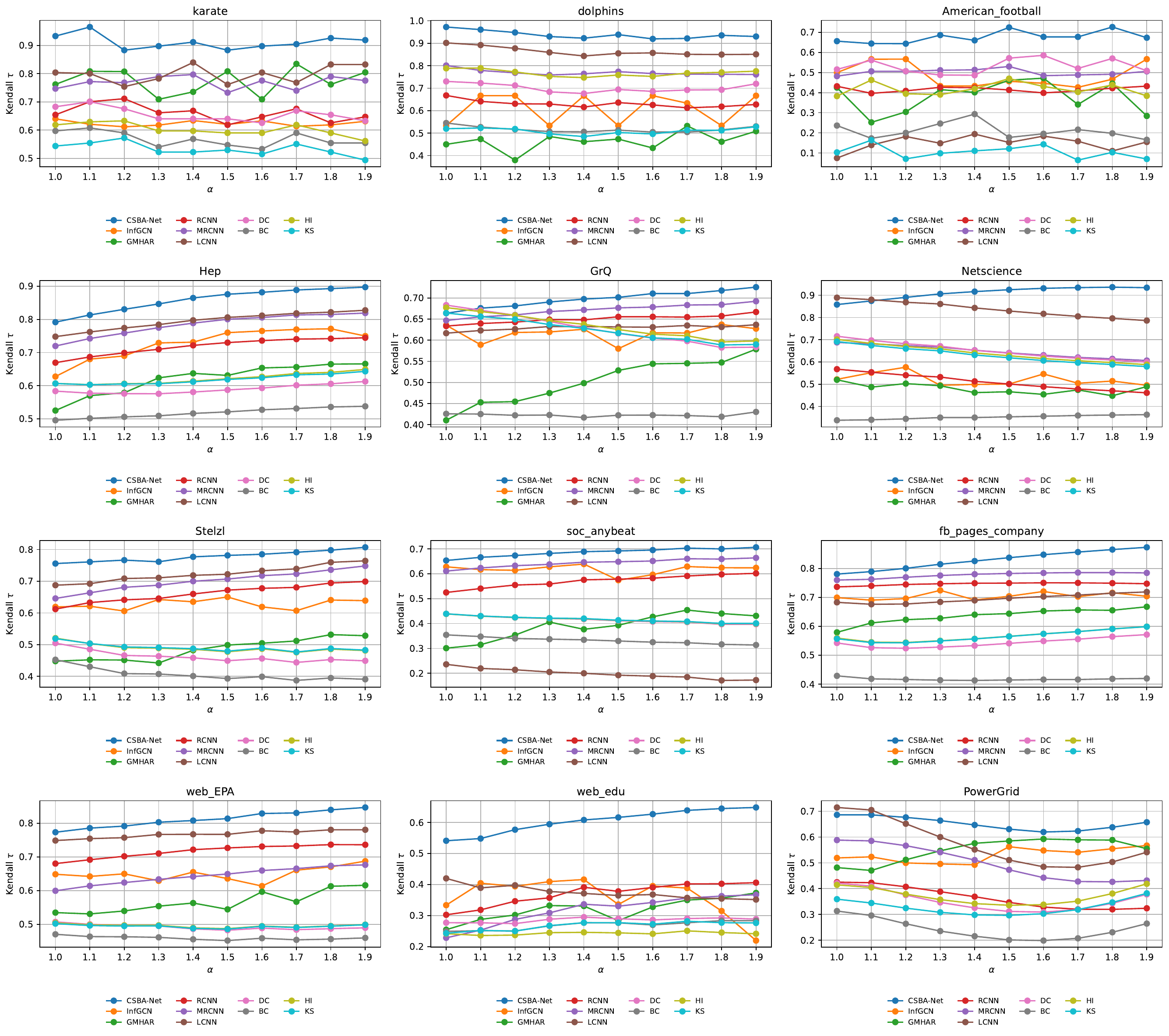}
\caption{Ranking performance comparison on complex networks: Kendall’s $\tau$ versus the infection parameter $\alpha \in [1,1.9]$. Higher $\tau$ indicates better agreement with SIR spreading influence.}
    \label{fig:knd_tau_part1}
\end{figure*}

\begin{figure*}[htbp]
    \centering
    \includegraphics[width=0.9\textwidth]{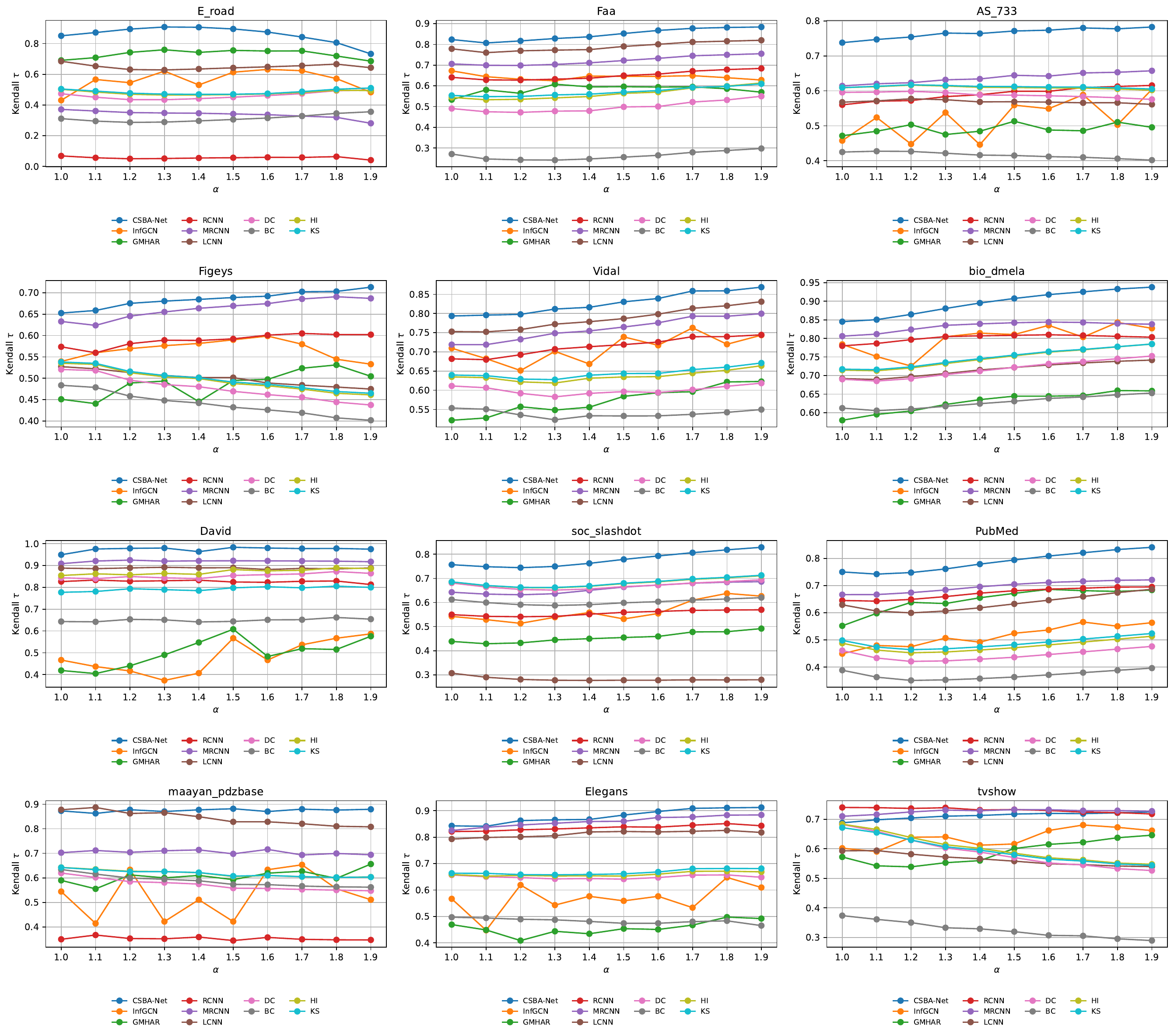}
\caption{ 
Ranking performance comparison on complex networks: Kendall’s $\tau$ versus the infection parameter $\alpha \in [1,1.9]$. Higher $\tau$ indicates better agreement with SIR spreading influence.
    %Performance comparison of ranking methods on complex networks. Each subplot shows Kendall’s $\tau$ correlation coefficient versus the infection parameter $\alpha$. The horizontal axis represents $\alpha \in [1, 1.9]$ while the vertical axis represents Kendall's $\tau$. Higher values indicate better agreement between the ranking method and the spreading influence obtained from SIR simulations.
    }
    \label{fig:knd_tau_part2}
\end{figure*}

We conduct this analysis on all 24 benchmark real-world networks, and the obtained results are reported in \textit{Figures~\ref{fig:knd_tau_part1}} and\textit{~\ref{fig:knd_tau_part2}}.% (for more details about the results obtained on Barabási–Albert  networks, refer to \textit{Figure \ref{fig:knd_tau_ba}} in \textit{Appendix}).% reporting Kendall’s~$\tau$ as a function of $\alpha$ for all compared methods.

%This analysis examines how changes in the diffusion regime used to generate the ground-truth influence scores affect the ranking performance of CSBA-Net.

%This allows us to determine whether the quality of the predicted influential-node ranking remains stable when the underlying diffusion process becomes progressively more intense. In addition to the average performance at each diffusion level, we examine the variation of Kendall’s $\tau$ across different $\alpha$ values to assess the stability of the model.

%This analysis provides a complementary perspective to the structural and cross-domain evaluations in Section 6.2. While those experiments assess how CSBA-Net behaves across different network structures and domains, the present experiment focuses on the effect of the diffusion condition itself, thereby evaluating the sensitivity and stability of the learned ranking with respect to the SIR diffusion intensity.

\vspace{0.15cm}
\noindent\textit{Performance trend with $\alpha$.%Global ranking and monotonic trend.
} CSBA-Net achieves the highest Kendall’s~$\tau$ on 19 of 24~networks at $\alpha=1.0$ and on 23 of 24 for all $\alpha\geq1.2$, without $\alpha$-specific tuning. The mean Kendall’s~$\tau$ across the 24 networks increases from $0.776$ at $\alpha=1.0$ to $0.822$ at $\alpha=1.8$, then stabilizes at $0.821$ at $\alpha=1.9$. This improvement suggests that increasing infection probability makes influential-node rankings more discriminative, benefiting methods that capture structural spreading capacity. In fact, CSBA-Net’s multi-scale representations, encoded in $D^u$, $B^u$, and $H^u$, enable it to capture structural properties relevant to diffusion across multiple propagation scales.

\vspace{0.15cm}
\noindent\textit{Behavior on sparse networks.%Baseline instability on sparse networks.
} 
Baseline methods show greater instability on sparse and irregular networks. On PowerGrid ($\langle k\rangle=2.67$), RCNN and LCNN exhibit substantial declines in Kendall’s~$\tau$, from $0.425$ to $0.320$ and from $0.715$ to $0.485$, respectively, whereas CSBA-Net remains within $[0.620,0.686]$. On E\_Road ($\langle k\rangle=2.41$), the performance of CSBA-Net varies from $0.908$ at $\alpha=1.3$ to $0.733$ at $\alpha=1.9$, reflecting the challenges of ranking nodes in highly sparse networks under strong diffusion. Nevertheless, it outperforms RCNN, whose $\tau$ remains below $0.068$, highlighting the robustness of the proposed structural representations.

\vspace{0.15cm}
% \noindent\textit{Stability across $\alpha$.%Quantitative stability across network categories.

\noindent\textit{Stability across $\alpha$.} Based on the standard deviation of Kendall’s~$\tau$ across $\alpha$, we see that CSBA-Net remains stable, with a mean standard deviation of $0.024$ across the 24~networks and a maximum of $0.052$ on E\_Road. LCNN and M-RCNN reach higher deviations of $0.084$ and $0.064$, respectively, on PowerGrid. Across network categories, CSBA-Net exhibits similar variability in social ($0.022$), biological ($0.021$), and knowledge ($0.025$) networks, with slightly higher variation in technological networks ($0.029$), reflecting the difficulty posed by sparse, geographically constrained infrastructure graphs such as PowerGrid and E\_Road \cite{Lu2016vital, kitsak2010identification}.

\vspace{0.15cm}
\noindent\textit{Saturation at high diffusion intensity.} CSBA-Net’s mean Kendall’s~$\tau$ increases with $\alpha$, peaking at $\alpha=1.8$ and remaining nearly stable over $\alpha\in[1.5,1.9]$, with a variation below $0.013$. This plateau suggests that ranking quality saturates under moderate-to-high diffusion, where structural spreading capacity becomes more informative. Performance is slightly lower near $\alpha=1.0$, where stochastic effects around the epidemic threshold introduce greater variability. Nevertheless, the model outperforms most baselines, which suggests that the $T=1{,}000$ independent SIR simulations provide sufficiently stable ground-truth labels for robust regression.

Overall, the results show that the proposed framework remains robust across diffusion intensities and does not rely on a specific $\alpha$. Its stable ranking performance, including on sparse networks, supports the ability of its multi-scale structural representations to capture topology-driven properties relevant to spreading potential.

\section{Conclusions and Future Work} 

In this paper, we introduced CSBA-Net (\textit{Cross-Scale Structural BiGRU Attention Network}), a novel deep learning framework for influential node identification that addresses key limitations of existing approaches, including single-scale structural representations, binary formulations, and limited modeling of dependencies across different network scales. CSBA-Net integrates complementary micro-, meso-, and macro-level representations that capture local connectivity, inter-community bridging, and global reachability, respectively. These representations are processed independently by CNN branches and subsequently fused through a BiGRU and a multi-head self-attention mechanism to model dependencies across structural scales. In addition, the problem is formulated as a continuous regression task using SIR-based influence scores, enabling more fine-grained estimation of node influence. 

Extensive experiments on 24 real-world networks from four domains, together with 12 synthetic Barab\'{a}si--Albert networks, demonstrate the effectiveness and generalization of the proposed framework. CSBA-Net consistently outperforms centrality-based, CNN-based, and GNN-based baselines across Kendall's $\tau$, Spearman's $\rho$, NDCG@10, and NDCG@20, achieving average scores of $\bar{\tau}=0.802$, $\bar{\rho}=0.943$, $\text{NDCG@10}=0.898$, and $\text{NDCG@20}=0.916$. It ranks first in Spearman's $\rho$ on all 24 real-world networks and in Kendall's $\tau$ on 23 of them. Notably, the model is trained only on a single synthetic BA network and transferred directly to real-world networks without retraining, demonstrating strong cross-domain generalization.

The largest gains are observed on structurally challenging networks, particularly sparse and heterogeneous graphs, where conventional methods degrade more substantially. Ablation results confirm the importance of the proposed components, with BiGRU-based inter-scale fusion providing the largest contribution, followed by the multi-scale representation, structured neighborhood selection, and multi-head attention. Sensitivity and robustness analyses further show that performance remains stable across diffusion conditions, network domains, and network sizes. All these results support CSBA-Net as a generalizable framework that combines multi-scale structural encoding, inter-scale dependency modeling, and continuous influence estimation within a unified architecture.

\vspace{0.15cm}
\noindent\textit{Limitations and Future Work.} Despite these results, several limitations remain. First, CSBA-Net is currently designed for static, unweighted, and undirected networks. Extending the framework to temporal, directed, and weighted graphs would enable it to model evolving structures and weighted interactions in a broader range of real-world applications. Second, the computational cost of the current pipeline may limit its application to very large networks, particularly due to centrality computations such as betweenness, harmonic, and eigenvector centrality. Future work should therefore investigate scalable approximations and sampling-based strategies to reduce the cost of structural feature extraction and enable its application to larger graphs. Third, the framework relies exclusively on topological information and does not incorporate node attributes such as semantic representations or behavioral information. Integrating these signals with the proposed multi-scale structural representations could improve influence estimation, particularly in social and knowledge networks. Finally, extending CSBA-Net to temporal and attributed settings raises additional challenges in dynamically updating structural representations and modeling the evolution of influence over time. %Addressing these issues, together with improving computational scalability and supporting richer graph structures, constitutes a promising direction toward more expressive, efficient, and generalisable influential-node identification frameworks for complex real-world networks.

\bibliographystyle{abbrvnat}
\bibliography{reference.bib}

\end{document}